\pdfoutput=1
\documentclass{aa}  

\usepackage{color}
\usepackage{graphicx}
\usepackage{amsmath}
\usepackage{mathrsfs}
\usepackage{txfonts}
\usepackage{natbib,twoopt}
\usepackage[breaklinks=true]{hyperref} 
\bibpunct{(}{)}{;}{a}{}{,}             
\makeatletter
  \newcommandtwoopt{\citeads}[3][][]{\href{http://adsabs.harvard.edu/abs/#3}%
    {\def\hyper@linkstart##1##2{}%
     \let\hyper@linkend\@empty\citealp[#1][#2]{#3}}}
  \newcommandtwoopt{\citepads}[3][][]{\href{http://adsabs.harvard.edu/abs/#3}%
    {\def\hyper@linkstart##1##2{}%
     \let\hyper@linkend\@empty\citep[#1][#2]{#3}}}
  \newcommandtwoopt{\citetads}[3][][]{\href{http://adsabs.harvard.edu/abs/#3}%
    {\def\hyper@linkstart##1##2{}%
     \let\hyper@linkend\@empty\citet[#1][#2]{#3}}}
  \newcommandtwoopt{\citeyearads}[3][][]%
    {\href{http://adsabs.harvard.edu/abs/#3}
    {\def\hyper@linkstart##1##2{}%
     \let\hyper@linkend\@empty\citeyear[#1][#2]{#3}}}
\makeatother
\definecolor{mygreen}{RGB}{0, 139, 69}

\DeclareSymbolFont{mymath}{T1}{ybv}{m}{it}
\SetSymbolFont{mymath}{normal}{T1}{ybv}{m}{it}
\DeclareSymbolFontAlphabet{\mathnormal}{mymath}
\DeclareMathSymbol{\vel}{\mathalpha}{mymath}{`v}

\begin{document} 

   \title{High-resolution three-dimensional simulations of gas removal from 
     ultrafaint dwarf galaxies}

   \subtitle{I. Stellar feedback}

   \author{Donatella Romano
          \inst{1}
          \fnmsep\thanks{donatella.romano@inaf.it},
          Francesco Calura
          \inst{1},
          Annibale D'Ercole
          \inst{1},
          \and
          C.~Gareth Few
          \inst{2,3}
          }

   \institute{INAF, Astrophysics and Space Science Observatory,
              via Gobetti 93/3, I-40129, Bologna, Italy
         \and E.A. Milne Centre for Astrophysics, University of Hull, 
              Cottingham Road, Kingston Upon Hull, HU6 7RX, UK
         \and Joint Institute for Nuclear Astrophysics--Center for the 
              Evolution of the Elements (JINA--CEE)
             \\
             }

   \date{Received 21 February 2019 / Accepted 4 September 2019}

   \titlerunning{High-resolution simulations of gas removal from UFDs}
   \authorrunning{Romano et al.}

  \abstract
   {The faintest Local Group galaxies found lurking in and around the Milky Way 
     halo provide a unique test bed for theories of structure formation and 
     evolution on small scales. Deep Subaru and \emph{Hubble Space Telescope} 
     photometry demonstrates that their stellar populations are old, and that 
     the star formation activity did not last longer than 2~Gyr in these 
     systems. A few mechanisms that may lead to such a rapid quenching have 
     been investigated by means of hydrodynamic simulations, without providing 
     any final assessment so far.}
   {This is the first in a series of papers aimed at analysing the roles of 
     stellar feedback, ram pressure stripping, host-satellite tidal 
     interactions and reionization in cleaning the lowest-mass Milky Way 
     companions of their cold gas, by using high-resolution, three-dimensional 
     hydrodynamic simulations.}
   {We simulate an isolated ultrafaint dwarf galaxy loosely modeled after 
     Bo\"otes~I, and examine whether or not stellar feedback alone could drive 
     a substantial fraction of the ambient gas out from the shallow potential 
     well.}
   {In contrast to simple analytical estimates, but in agreement with previous 
     hydrodynamical studies, we find that most of the cold gas reservoir is 
     retained. Conversely, a significant fraction of the metal-enriched stellar 
     ejecta crosses the boundaries of the computational box with velocities 
     exceeding the local escape velocity and is, thus, likely lost from the 
     system.}
   {Although the total energy output from multiple supernova explosions exceeds 
     the binding energy of the gas, no galactic-scale outflow develops in our 
     simulations and as such, most of the ambient medium remains trapped within 
     the weak potential well of the model galaxy. It seems thus unavoidable 
     that, in order to explain the dearth of gas in ultrafaint dwarf galaxies, 
     we will have to resort to environmental effects. This will be the subject 
     of a forthcoming paper.}


   \maketitle
%

\section{Introduction}
\label{sec:intro}

   In the $\Lambda$~cold dark matter concordance cosmology, large haloes form 
   from the merging and accretion of small building blocks 
   \citep[][]{1978MNRAS.183..341W}. A large number of subhaloes, however, are 
   predicted to survive this digestion process and inhabit today's main galaxy 
   haloes. The mismatch between theoretical expectations and observed Galactic 
   satellites gave rise to a highly-debated issue, the so-called `missing 
   satellite problem' \citep[MSP;][]{1999ApJ...516..530K, 1999ApJ...524L..19M}. 
   In the past dozen years, though, the number of known Milky Way (MW) 
   companions has increased apace, thanks to the commitment of hundreds of 
   scientists around the world to deep large-area sky imaging surveys, such as 
   the Sloan Digital Sky Survey \citep[][]{2000AJ....120.1579Y}, the VST ATLAS 
   Survey \citep[][]{2015MNRAS.451.4238S}, the Dark Energy Survey 
   \citep[][]{2016MNRAS.460.1270D}, the Pan-STARRS1 3\,$\pi$ Survey 
   \citep[][]{2016arXiv161205560C}, and the Hyper Suprime-Cam Subaru Strategic 
   Program Survey \citep[][]{2018PASJ...70S...4A}. A new class of galaxy has 
   been discovered, one made of extremely faint, dark-matter dominated, 
   scarcely evolved stellar systems, that have been given the name of 
   ultrafaint dwarf galaxies \citep[UFDs; see, e.g.,][]{2005AJ....129.2692W,   
   2006ApJ...647L.111B, 2007ApJ...654..897B, 2014MNRAS.441.2124B,   
   2015ApJ...807...50B, 2015ApJ...813...44L, 2016ApJ...832...21H,   
   2018PASJ...70S..18H}. Deep \emph{Hubble Space Telescope} observations of six 
   UFDs, reaching below the main-sequence turnoff, are best fit by ancient 
   stellar populations \citep[11.6~Gyr old,][]{2014ApJ...796...91B}. From deep 
   images obtained with the Suprime-Cam on the Subaru Telescope, 
   \citet{2012ApJ...744...96O} conclude that most stars in Bo\"otes~I, one of 
   the best studied UFDs, are consistent with a single-epoch, short burst of 
   star formation.

   All of this has dramatically impacted our understanding of the way the halo 
   of our Galaxy came into being. The continuous discovery of more and more 
   UFDs, even at the detectability limits of the surveys \citep[see][their 
     figure~9]{2016MNRAS.463..712T}, alleviates considerably the need to resort 
   to alternative dark matter scenarios to address the MSP \citep[see][and 
     references therein]{2008ApJ...688..277T}, and moves the focus on studies 
   aimed at establishing how star formation and chemical enrichment proceed in 
   smaller and smaller dark matter haloes.

   Semi-analytical and pure chemical evolution models have been used in the 
   first place to follow the evolution of different elements in the 
   interstellar medium (ISM) of systems with structural properties resembling 
   those of UFDs. By introducing simple, heuristic (yet physically-motivated) 
   recipes to treat complex processes, such as accretion, cooling, star 
   formation and radiative feedback, these models can efficiently (in terms of 
   computational costs) explore a wide range of parameter space. It is 
   concluded that UFDs formed stars very inefficiently -- they converted less 
   than 1--3 per cent of their baryons into stars. However, there is no 
   consensus about the dominant mechanism that truncates star formation, with 
   either reionization \citep[][]{2009MNRAS.395L...6S}, galactic winds 
   \citep[][]{2014MNRAS.441.2815V} or tidal stripping 
   \citep[][]{2015MNRAS.446.4220R} being preferred. Chemical evolution models 
   have been very successful in explaining many observed properties of galaxies 
   but, when moving to stellar systems with lower and lower dynamical masses, 
   the parameterizations adopted to treat the thermal feedback from stars, the 
   conditions imposed on the onset of galactic-scale outflows, and the assumed 
   mass loading factors introduce severe degeneracies in the proposed 
   solutions. Furthermore, there are important limitations in the treatment of 
   spatial inhomogeneities. That is why pure chemical evolution models can not 
   put firm constraints on the physical processes regulating the evolution of 
   the lowest mass systems \citep[see, e.g., a discussion 
     in][]{2015MNRAS.446.4220R}. The natural outcome is to turn to 
   hydrodynamical simulations. 

   There are not, actually, that many such studies devoted to UFDs in the 
   literature. \citet[][]{2015ApJ...807..154B} adopted the three-dimensional 
   hydro/ionization code \emph{Fyris Alpha} \citep[][]{2010Ap&SS.327..173S} to 
   track the response of adiabatic and cooling models, with clumpy or smooth 
   gas distributions, to a single supernova (SN) explosion, occurring either at 
   the centre or off-centre of isolated UFD-sized haloes. Building on this 
   work, \citet[][]{2015ApJ...799L..21W} evolved $M_{\rm{vir}}$~= 10$^7$~M$_\odot$ 
   systems in isolation with extended star formation to fit the stellar 
   metallicity distribution functions and [$\alpha$/Fe] ratios observed in six 
   UFDs. Their main conclusion is that the UFDs form in low-mass haloes, rather 
   than being remnants of larger systems.

   High-resolution studies of the co-evolution of a UFD with a MW-like galaxy 
   over a Hubble time remain out of reach of present computational capabilities 
   \citep[][]{2018NatAs.tmp...32F}. \citet[][]{2017ApJ...848...85J} use a 
   customized version of the $N$-body/TreePM smoothed particle hydrodynamics 
   code GADGET \citep[][]{2001NewA....6...79S} to perform zoom-in simulations 
   of relatively isolated systems outside of the virial radius of a MW-like 
   host halo. This choice allows them to minimize the computational cost by 
   excluding processes such as tidal interactions and ram pressure 
   stripping\footnote{A remarkable attempt to consider the role of stripping on 
     the evolution of the lowest-mass MW satellites is made by 
     \citet[][]{2016ApJ...826..148E}. Although their $\sim$10~pc resolution 
     simulations can not be considered fully converged \citep[see][their 
       figure~7]{2016ApJ...826..148E}, these authors can conclude that other 
     mechanisms besides stellar feedback and ram pressure stripping must be at 
     play in order to reconcile the theoretical quenching timescales with those 
     deduced from observations.}. They find that haloes with virial mass 
   $M_{\rm{vir}}$~$\le$~2~$\times$~10$^9$~M$_\odot$ form the bulk of their stellar 
   populations before reionization, and confirm that the combined effect of 
   reionization and SN feedback is responsible for quenching the star formation 
   in these systems \citep[see also][]{2015MNRAS.453.1305W,2016MNRAS.456...85S}.
   They also find that accretion and mergers may play an important role in the 
   assembly history of UFDs. More recently, \citet[][]{2018MNRAS.475.4868C} 
   study the formation and evolution of UFDs in the context of the 
   cosmological, Adaptive Mesh Refinement (AMR) radiation hydrodynamics 
   simulations discussed by \citet[][]{2012MNRAS.427..311W}. The maximum 
   comoving resolution is 1~pc, at the forefront of current cosmological 
   simulations. Although some important observational properties of Local Group 
   UFDs are recovered, some limitations remain. For instance, the authors 
   acknowledge that the stellar metallicity distribution functions in the 
   simulation are too narrow and metal-rich.

   This paper is the first in a series aimed at analysing the roles of internal 
   processes, such as stellar feedback, and external ones, such as ram pressure 
   stripping, host-satellite tidal interactions and reionization, in the 
   evolution of the lowest-mass MW satellites, paying particular attention to 
   the way they lose their cold ($T < 10^4$~K) gas. We perform 
   three-dimensional adiabatic and radiative simulations for an isolated system 
   loosely resembling the UFD Bo\"otes~I, and examine whether or not stellar 
   feedback alone is able to drive a significant gas fraction out from the 
   simulation volume. The paper is organized as follows. In 
   Section~\ref{sec:setup}, we describe the adopted numerical set-up. In 
   Section~\ref{sec:res}, we present the results from our fiducial 
   high-resolution runs, with and without radiative cooling (a suite of lower 
   resolution simulations is analysed in Appendix~\ref{sec:AppA}). We 
   demonstrate that the neutral ambient medium is mostly unaffected by the SN 
   events, even when their effect is maximized in the simulation. However, the 
   newly-produced metals efficiently escape the galaxy potential well. In 
   Section~\ref{sec:disc}, we discuss our results and compare them to the 
   expectations from analytic computations, as well as to previous work in the 
   literature. Finally, in Section~\ref{sec:conc} we draw our conclusions.

\section{Numerical set-up}
\label{sec:setup}

   We have run our simulations with a customized version of the AMR code 
   \textsc{ramses} \citep[][]{2002A&A...385..337T} that solves the Euler 
   equations of gravitohydrodynamics with a second-order, unsplit Godunov 
   scheme. The fluid follows the adiabatic equation of state for an ideal 
   mono-atomic gas with adiabatic index $\gamma = 5/3$.

   The initial configuration is designed to mimic the Bo\"otes~I UFD, for which 
   extensive observational work has been published elsewhere \citep[see][and 
     references therein]{2015MNRAS.446.4220R}, and foresees a non-rotating 
   distribution of gas and stars embedded in an isolated dark matter halo. The 
   theoretical initial baryonic and dark masses are $M_{\rm{gas}}$~= 6~$\times$ 
   10$^6$~M$_\odot$ and $M_{\rm{DM}}$~= 3.5~$\times$ 10$^7$~M$_\odot$ and follow, 
   respectively, a \citet[][]{1911MNRAS..71..460P} density profile, with 
   characteristic radius $a \simeq$ 200~pc, and a Burkert's 
   (\citeyear[][]{1995ApJ...447L..25B}) profile, with cut-off radius 
   $R_{\rm{c}} \simeq$1.2~kpc. The gas is assumed to follow a smooth, 
   single-phase distribution reflecting the one observed nowadays for 
   long-lived stars in Bo\"otes~I. The dark matter component is modeled as a 
   static external potential and added to the solution of the Poisson equation; 
   owing to the dark matter dominance, the self-gravity of the gas is 
   neglected, for simplicity. The initial pressure profile is set by solving 
   the hydrostatic equilibrium\footnote{We checked that the hydrostatic 
     equilibrium condition is satisfied by evolving adiabatically the 
     unperturbed system at low resolution for 30~Myr.} equation. The initial 
   temperature profile is pretty flat, and ranges from $\sim$4400~K in the 
   central region to $\sim$3900~K in the outer zones.

   A population of coeval stars with mass $M_{\rm{stars}}$~= 10$^5$~M$_\odot$ is 
   set in place at the beginning of the simulation. In this work, we consider 
   an instantaneous star formation episode. This choice is made in order to 
   maximize the effects of stellar feedback and to obtain the highest synergy 
   among the stellar-wind and SN-driven bubbles. In the framework of the 
   integrated galactic initial mass function (IGIMF) theory \citep[][and 
     references therein]{2017A&A...607A.126Y,2018A&A...620A..39J}, a star 
   formation lasting a few tens to hundreds of Myr, or longer, would imply that 
   the stars form in small embedded clusters, thus leading to far fewer SN 
   events than expected for a canonical stellar initial mass function (IMF) as 
   the one adopted in this work \citep[][]{2001MNRAS.322..231K}. Assuming a 
   canonical stellar IMF, with lower and upper mass limits of 0.1 and 100 
   M$_\odot$, respectively, 650 of the stars are assigned initial masses in 
   excess of 8~M$_\odot$. These massive stars are grouped in ten OB associations 
   2\,$r_{\rm{OB}}$ wide (see next paragraph) scattered across the computational 
   volume, following the procedure outlined in \citet[][]{2015ApJ...814L..14C}. 
   We note that, because of our idealized set-up, some OB associations may fall 
   in regions of very low gas density (see previous paragraph). The implicit 
   assumption can be made that the OB associations formed in small overdense 
   regions -- a few spots with the right characteristics to collapse and form 
   stars. However, \citet{2019ApJ...872...16D} have recently shown that in 
   local dwarf galaxies the Schmidt law for star formation 
   \citep{1959ApJ...129..243S} does not seem to exhibit any threshold, as even 
   the lowest surface brightness regions show the presence of very young stars. 
   An even more extreme confirmation of this is represented by an utterly low 
   density ($n \sim$ 0.01~cm$^{-3}$) -- yet, star-forming -- galaxy recently 
   found in the Virgo cluster \citep{2018MNRAS.476.4565B}. All in all, since 
   UFDs are thought to represent the extreme limit of the galaxy formation 
   process \citep{2019arXiv190105465S}, the assumed gas and star distributions 
   for our model Bo\"otes~I galaxy may well be not quite so unreasonable. More 
   details about the sampling procedure and a discussion of the dependence of 
   the results on the particular set-up choice are deferred to 
   Appendix~\ref{sec:AppA}.

   Each association is allowed to inject mass and energy in its surroundings at 
   a constant rate for an uninterrupted period of 30~Myr (roughly corresponding 
   to the lifetime of a 8~M$_\odot$ star) through both stellar winds and SN 
   explosions. The energy input from discrete SNe can be reasonably 
   approximated as a continuous luminosity as long as the blast waves become 
   subsonic before cooling radiatively. This condition is usually met for SNe 
   exploding in a superbubble inflated by previous stellar wind activity 
   \citep[see][their section~III]{1988ApJ...324..776M}. After 
   \citet[][]{2014ApJS..212...14L}, during the pre-SN phase (lasting 3~Myr) an 
   OB association containing $N$ high-mass stars injects $N$\,(5~$\times$ 
   10$^{-8}$) M$_\odot$~yr$^{-1}$ and $N$\,(3~$\times$ 10$^{35}$) erg~s$^{-1}$, 
   while during the SN phase (i.e., from 3 to 30~Myr) these quantities rise to 
   $N$\,(4~$\times$ 10$^{-7}$) M$_\odot$~yr$^{-1}$ and $N$\,(7~$\times$ 10$^{35}$) 
   erg~s$^{-1}$. Mass and energy are spread on the volume occupied by the OB 
   association, that is $V_{\rm{OB}} =$~$\frac{4}{3}\,\pi\,r_{\rm{OB}}^3$, with 
   $r_{\rm{OB}} =$~4~pc for our high-resolution simulations. The massive star 
   feedback is modelled through thermal energy deposition, and no other 
   mechanism (such as, for instance, radiation pressure) is included. Unlike 
   other authors \citep[e.g.][]{2010MNRAS.405.1634S}, we do not include any 
   subgrid turbulence model.

   \begin{figure*}
   \centering
   \includegraphics[width=0.725\textwidth]{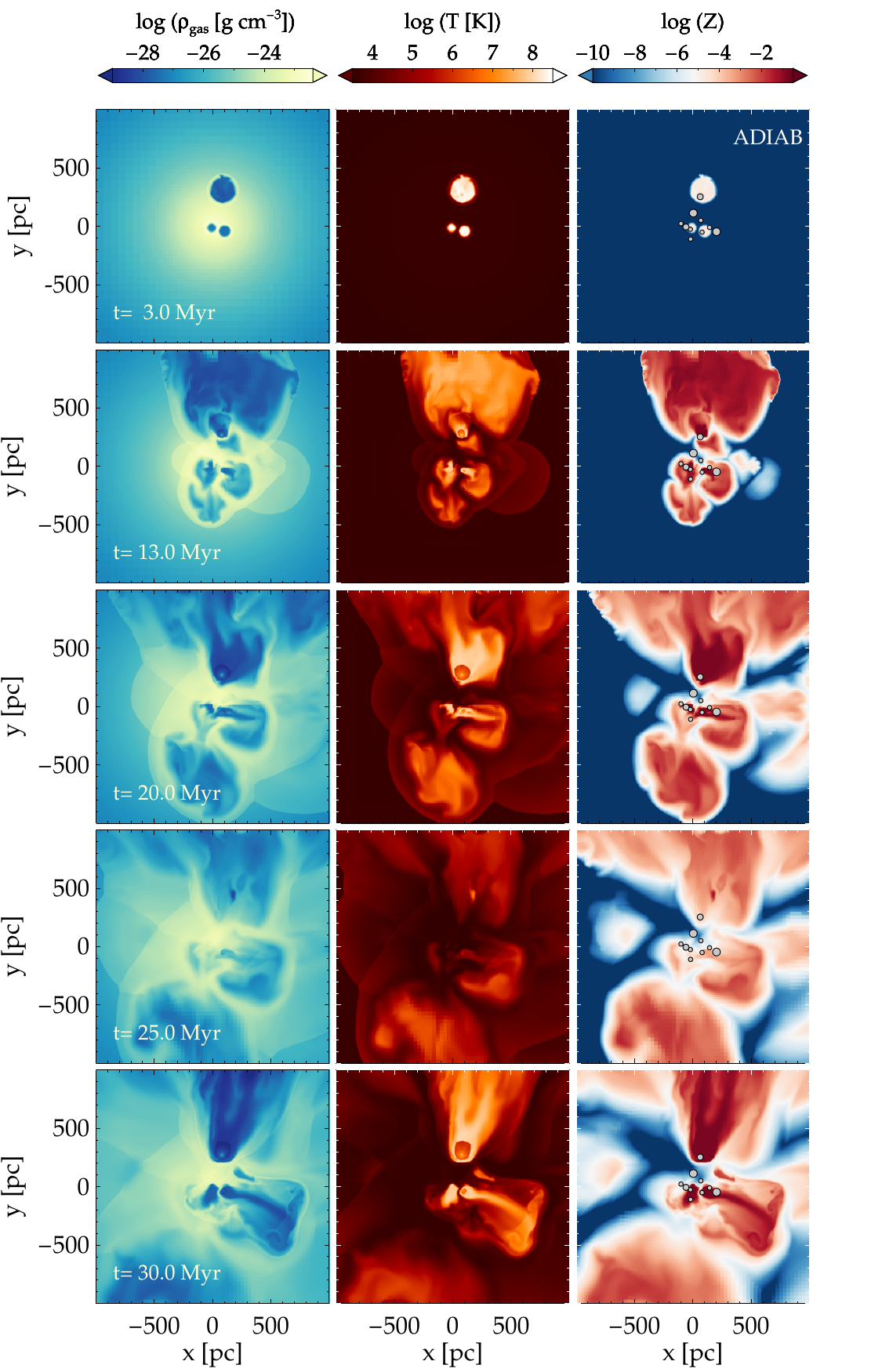}
   \caption{ Gas density (left), temperature (middle) and metallicity (right) 
     maps for the Bo\"otes~I-like galaxy, in the $z =$~0 plane, at five 
     representative times, $t$~= 3, 13, 20, 25, and 30~Myr. The projected 
     positions of the OB associations are displayed on the metallicity maps 
     (grey circles, with sizes proportional to the number of massive stars). 
     The snapshots refer to the galaxy simulated in the adiabatic limit at high 
     resolution.}
   \label{fig:mapdtz1}
   \end{figure*}

   \begin{figure*}
   \centering
   \includegraphics[width=0.725\textwidth]{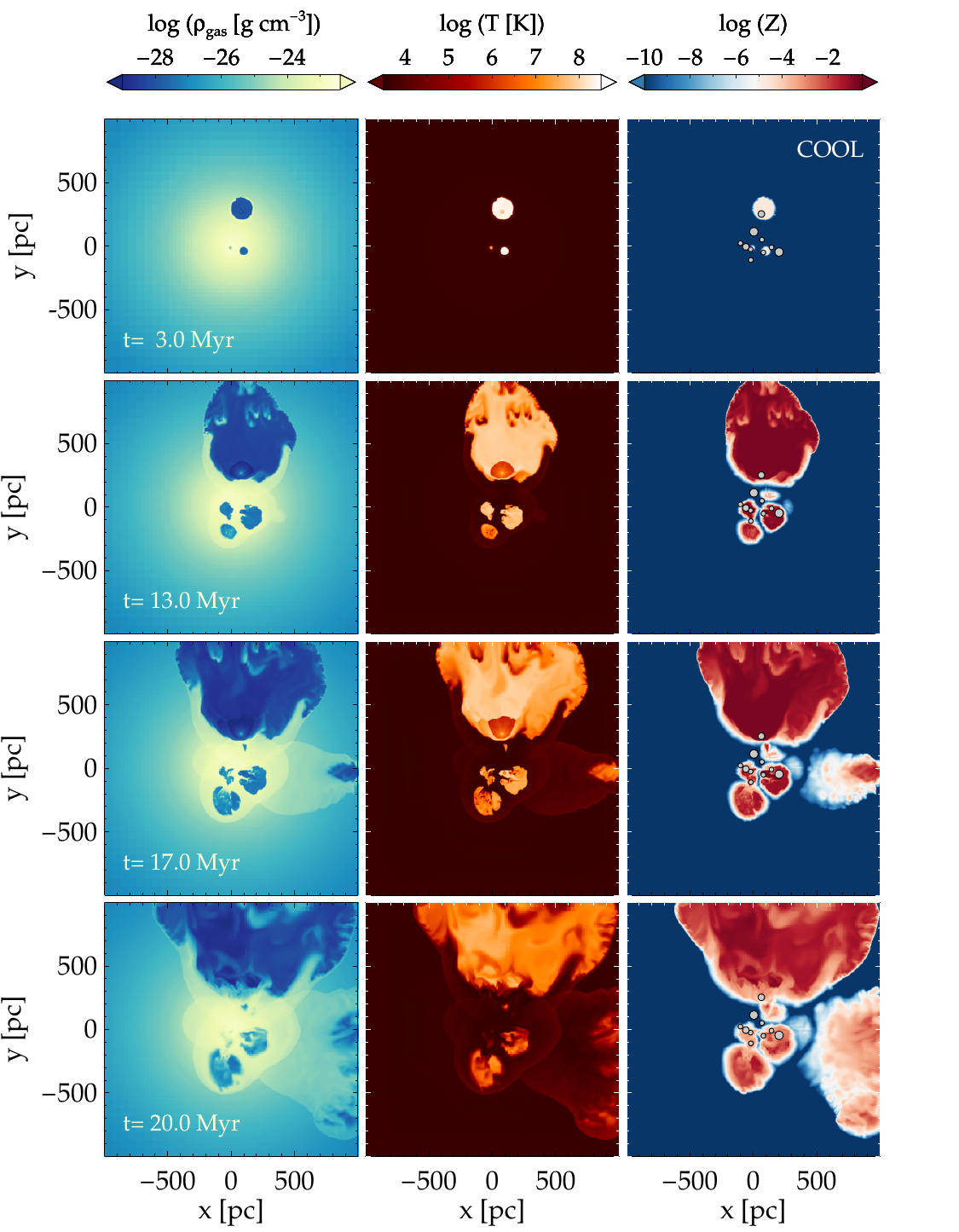}
   \caption{ Same as Fig.~\ref{fig:mapdtz1}, for $t$~= 3, 13, 17, and 20~Myr, 
     for the high-resolution simulation with radiative cooling.}
   \label{fig:mapdtz2}
   \end{figure*}

   \begin{figure*}
   \centering
   \includegraphics{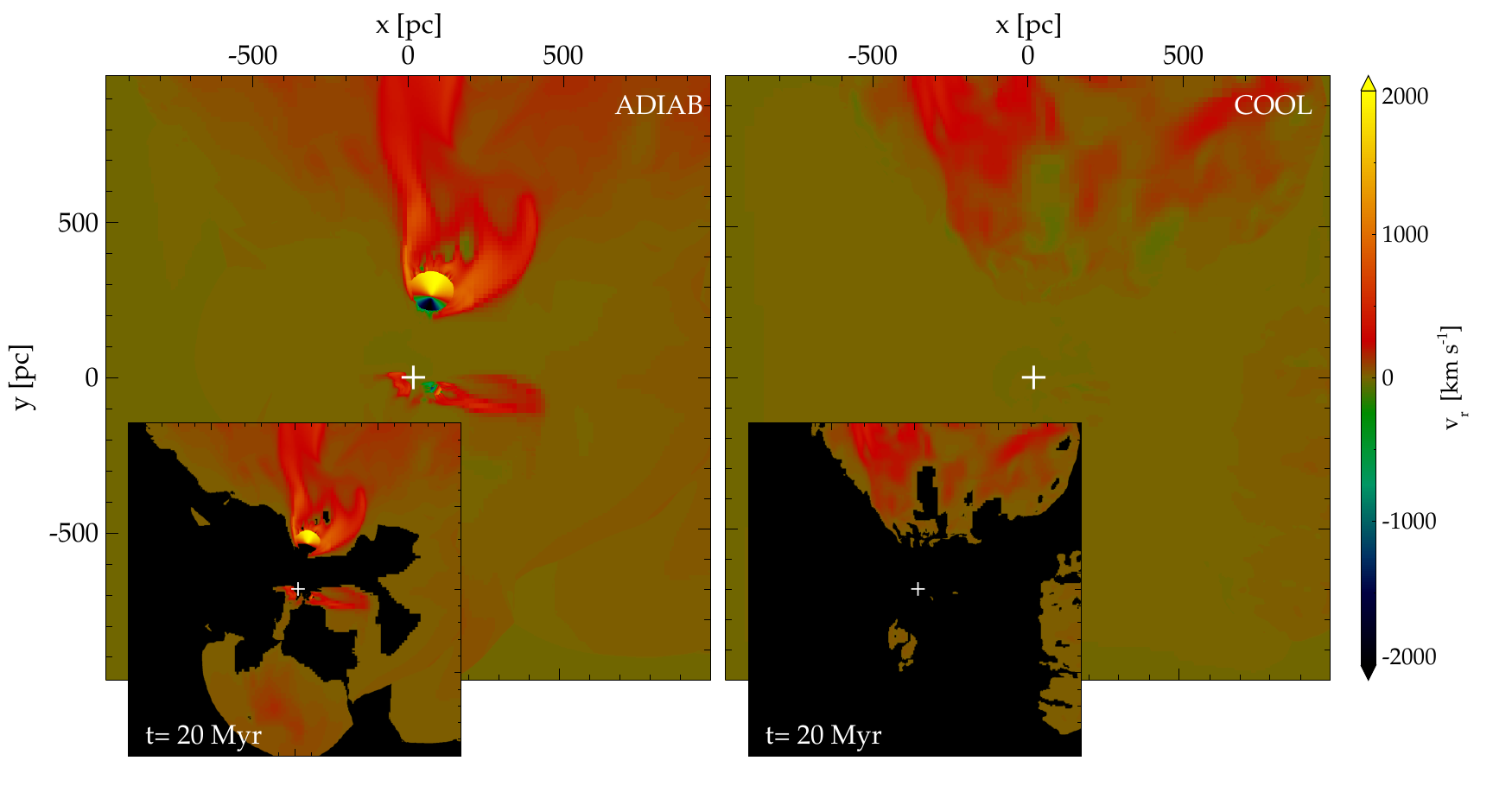}
   \caption{ Gas radial velocity field, in the $z =$~0 plane, at $t$~= 20~Myr. 
     The maps on the left refer to the galaxy simulated in the adiabatic limit, 
     while the maps on the right are for the run with radiative cooling. Cells 
     coloured black in the inset maps on the bottom left of each panel 
     highlight regions where the gas moves with velocities lower than the local 
     escape velocity.}
   \label{fig:mapvr20}
   \end{figure*}

   \begin{figure}
   \centering
   \includegraphics{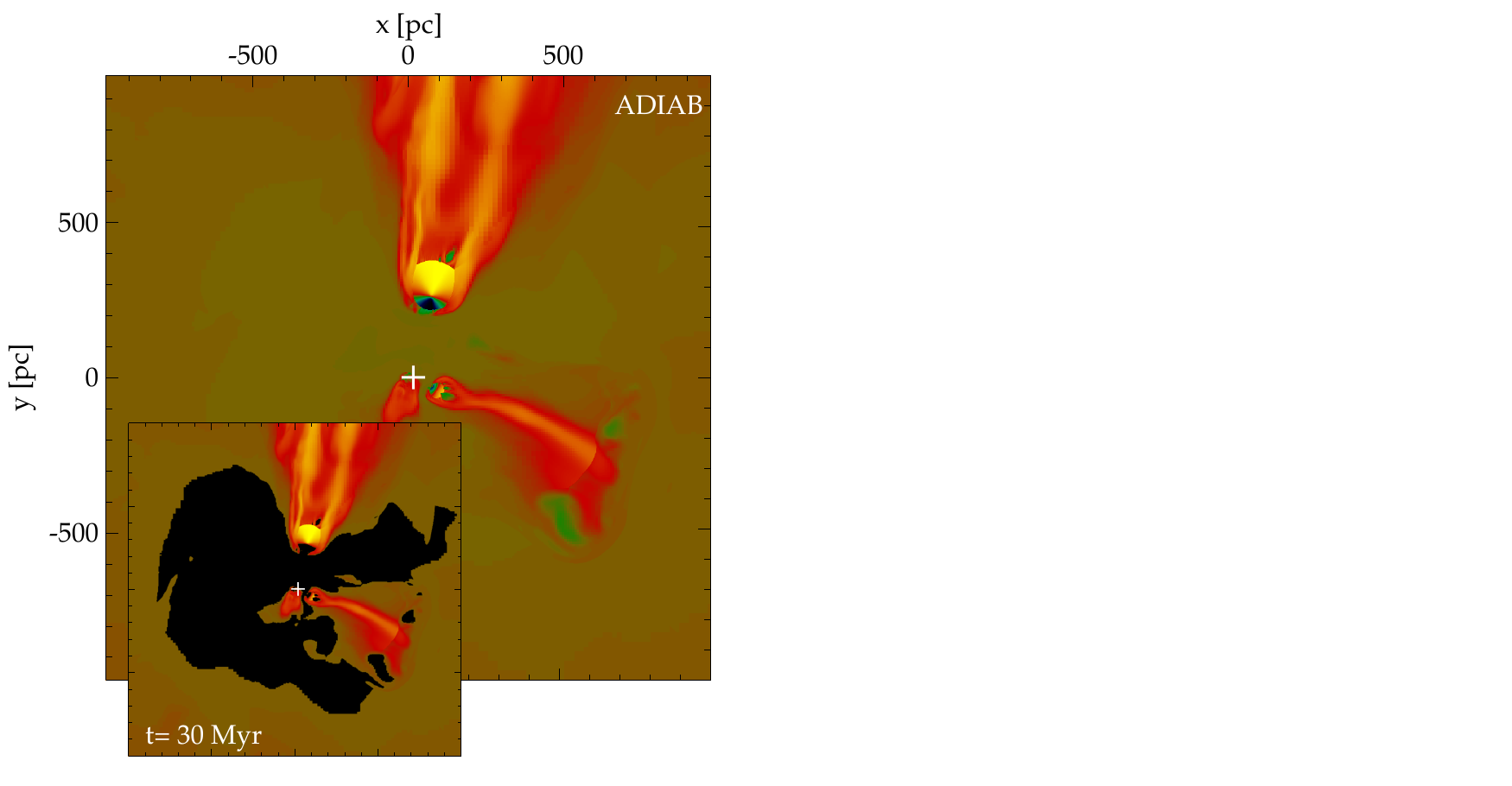}
   \caption{ Same as Fig.~\ref{fig:mapvr20}, for $t$~= 30~Myr, only for the 
     adiabatic simulation.}
   \label{fig:mapvr30}
   \end{figure}

   The computational box is $L$~= 2~kpc on a side, with a maximum refinement 
   level of $\ell_{\rm{max}} =$ 11, corresponding to a minimum cell size of 
   $\Delta x_{\rm{min}}$~= 0.95~pc (the lower resolution simulations discussed in 
   Appendix~\ref{sec:AppA} have $\ell_{\rm{max}} =$ 9 and $\Delta x_{\rm{min}}$~= 
   3.80~pc). The refinement strategy is geometry- and discontinuity-based. In 
   particular, at each timestep a number of cells at the highest refinement 
   level is set up to cover the regions occupied by the OB associations. This 
   assures that every OB association is adequately spatially resolved --stellar 
   ejecta and SN energy are added to the gas within a sphere that is four grid 
   cells in radius, which prevents the occurrence of square-shaped shock 
   fronts. With the introduction of a discontinuity-based criterion for 
   refinement, in addition, we make sure that the growing of dynamical 
   instabilities at the bubbles' fronts, as well as their interactions and 
   merging, are followed at the highest refinement level.

   According to \citet{2015ApJ...802...99K}, for a discrete SN event occurring 
   in a uniform, unmagnetized medium consistent convergence of the results is 
   obtained for $\Delta x_{\rm{min}}$, $r_{\rm{init}} < r_{\rm{sf}}/3$, where 
   $r_{\rm{init}}$ is the initial size of the SN remnant and $r_{\rm{sf}} =$ 21~pc 
   ($n_{\rm H}$/1 cm$^{-3}$)$^{-0.46}$ is its size at shell formation. In our 
   simulations, $n_{\rm H} = 0.76\,(\rho_{\rm gas}/m_{\rm p}) \lesssim$ 7~cm$^{-3}$ 
   and $r_{\rm{sf}} \gtrsim$ 9~pc everywhere, thus our choice of 
   $\Delta x_{\rm{min}} \simeq$ 1~pc for the high-resolution runs satisfies the 
   first convergence criterion. As for the second criterion, we have 
   $r_{\rm{OB}} \sim r_{\rm{sf}}/2$ at the centre (considering OB associations 
   rather than single SNe). Although this leads to evolutionary history and 
   internal profiles of the innermost bubbles that differ from the converged 
   solutions, the momentum and kinetic energy are expected to be not too far 
   from the correct values \citep[see section~4.2 in][]{2015ApJ...802...99K}. 
   Caution is urged, however, in that the study of \citet{2015ApJ...802...99K} 
   refers to discrete sources. A new study including the effects of the 
   preceding action of low-luminosity stellar winds would be very helpful to 
   clarify how this works for a continuous energy injection. It could be that 
   any initial numerical overcooling would be negligible once the densities 
   around the OB associations have dropped slightly due to the pre-SN feedback, 
   but this still needs to be quantified in detail.

   We use free outflow boundary conditions and create a passive scalar, 
   $Z = \rho_Z/\rho_{\rm{gas}}$, to trace the evolution of the metallicity of the 
   gas in each cell, starting from a primordial (zero) metallicity value. The 
   simulation outputs have a time resolution of 1~Myr.

   We run both adiabatic and cooling models. While the adiabatic simulations 
   run from 0 to 30~Myr, the cooling model is truncated at 20~Myr for 
   computational reasons\footnote{Our high-resolution adiabatic simulation 
     to 30~Myr costs on the order of 300\,000 CPU hours, while the radiative 
     one required 700\,000 CPU hours out to 20~Myr.}. We use \textsc{ramses} 
   build-in cooling rates, namely, for temperatures $T >$~10$^4$~K, gas cooling 
   follows the cooling function of \citet[][]{1993ApJS...88..253S} involving 
   hydrogen, helium, and metals. Below such temperature threshold, only 
   fine-structure metal cooling is considered by adopting the cooling function 
   of \citet[][]{1995ApJ...440..634R}. In accordance with the idealized nature 
   of our simulations, we set a temperature floor ($T_{\rm{min}}$~= 3900~K), 
   which mimics the effects of several heating mechanisms -- for instance, 
   photoionization and photoelectric heating from stars, and heating from the 
   galactic ultraviolet (UV) background radiation -- within the model galaxy.

\section{Results}
\label{sec:res}

   In this section, we discuss the results of high-resolution simulations with 
   650 SN precursors sorted in 10 OB associations placed at different radii. In 
   Appendix~\ref{sec:AppA}, we provide a general overview of the dependence of 
   the results on the particular configuration chosen for the OB associations. 
   To this aim, we make use of a suite of lower resolution simulations.

   \begin{figure*}
   \centering
   \includegraphics{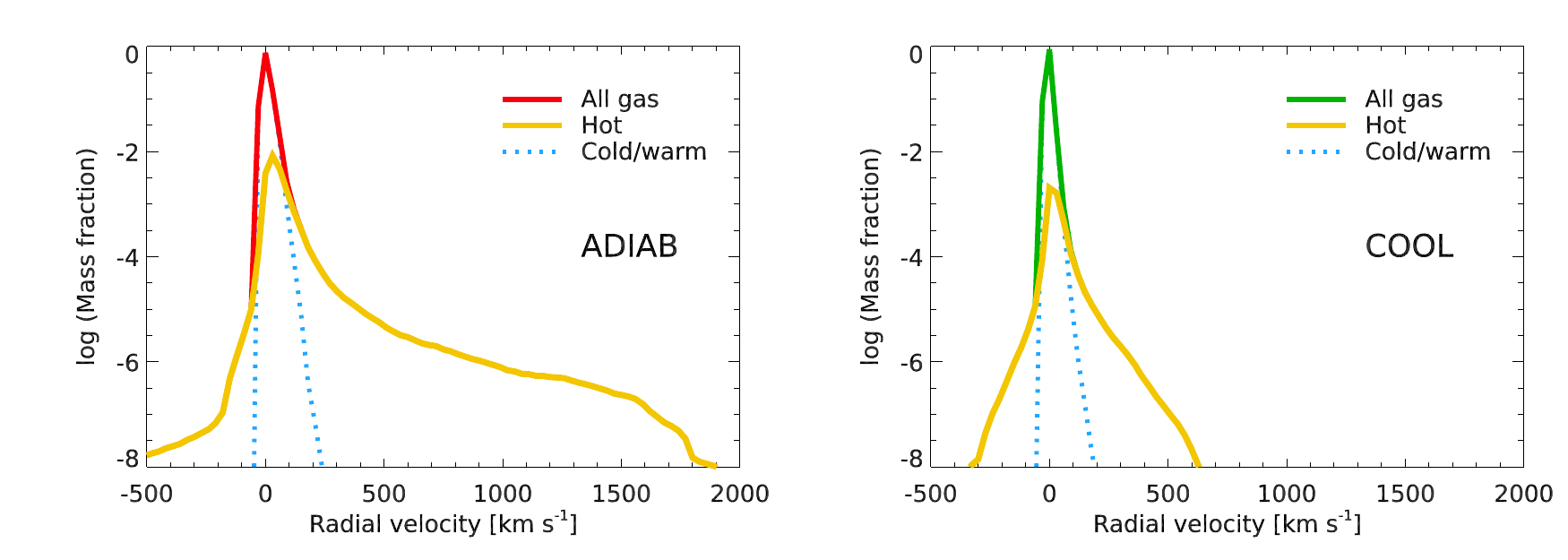}
   \includegraphics{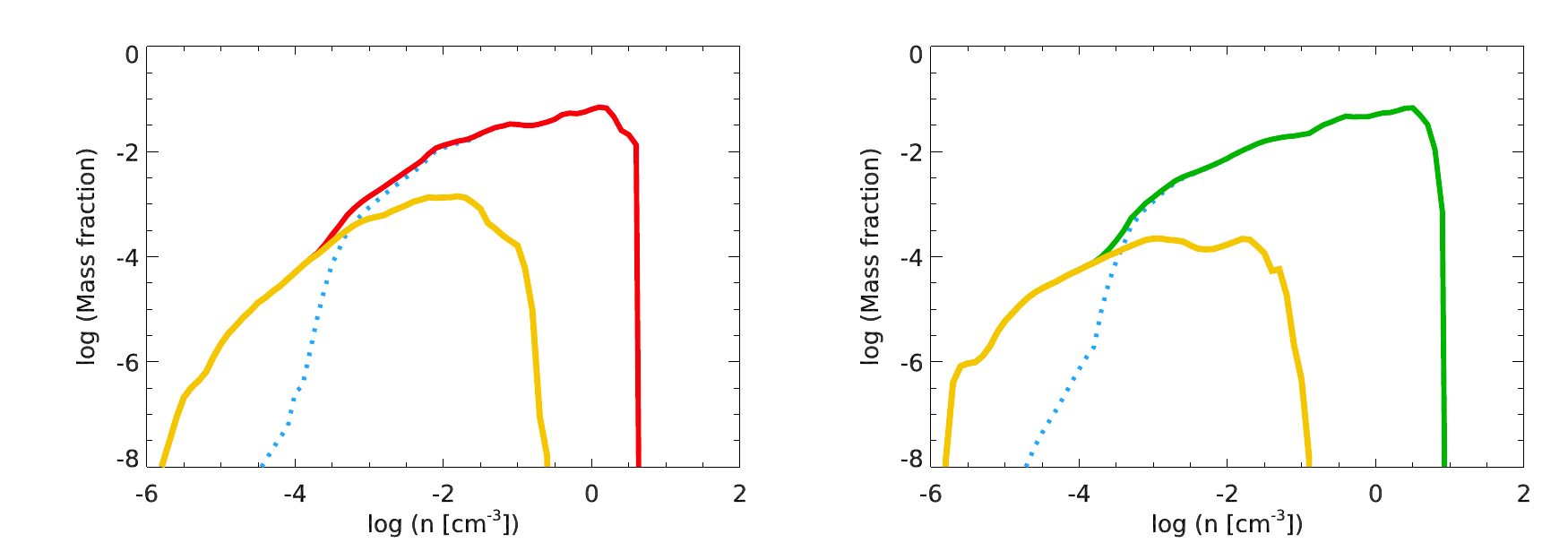}
   \includegraphics{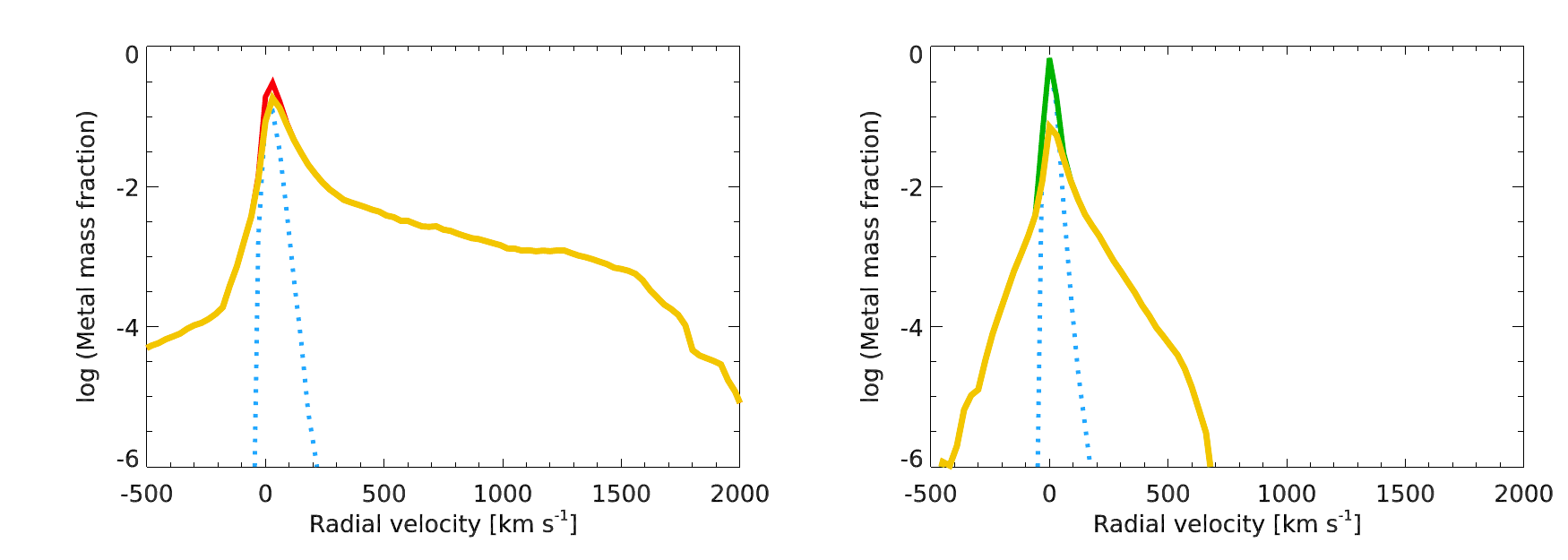}
   \caption{ Mass-weighted distributions of gas radial velocity (upper panels) 
     and density (middle panels) at $t$~= 20~Myr, for the runs with (right-hand 
     panels) and without radiative losses (left-hand panels). Also shown is the 
     distribution of metals in radial velocity bins at the same time for the 
     two runs (lower panels). The red (green) lines on the left (right) panels 
     show all the gas and metals in the simulation volume. The yellow solid and 
     blue dotted lines in all panels show the hot ($T \ge$ 10$^5$~K) and 
     cold/warm ($T <$ 10$^5$~K) components, respectively. The integrals of the 
     red and green lines are normalized to unity.}
   \label{fig:mwvd}
   \end{figure*}

   \begin{figure}
   \centering
   \includegraphics{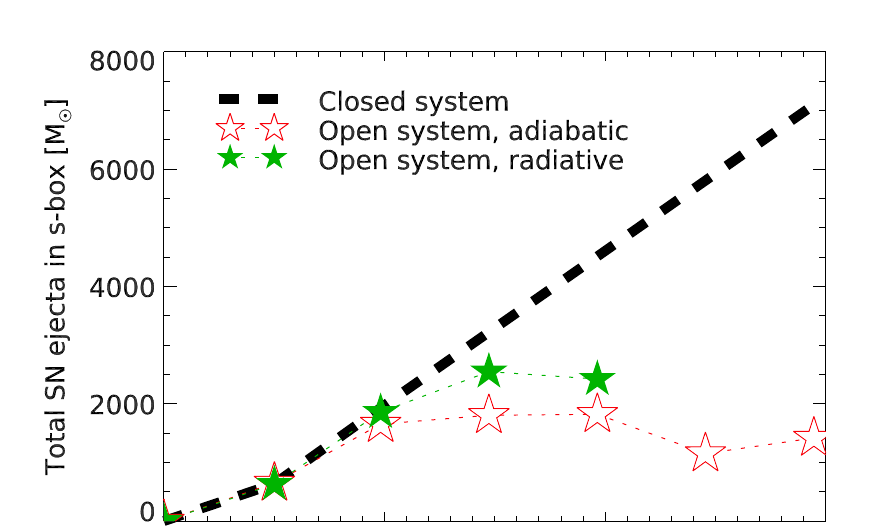}
   \includegraphics{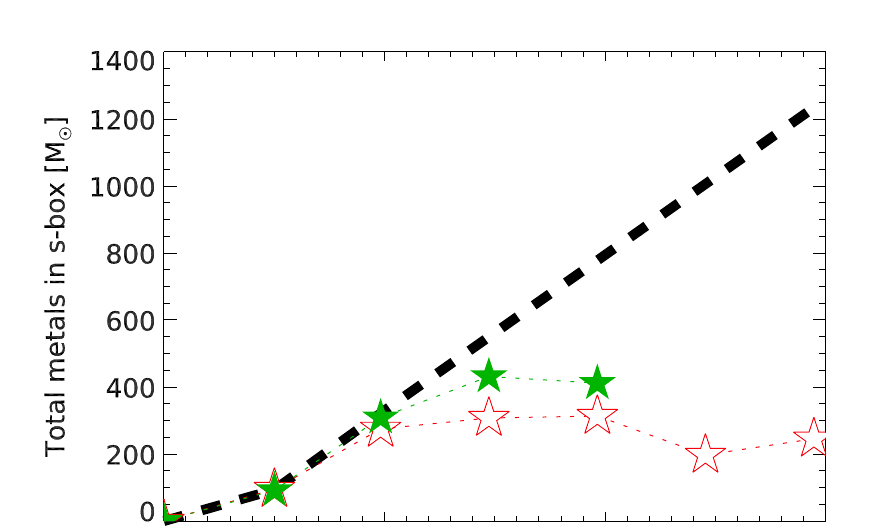}
   \includegraphics{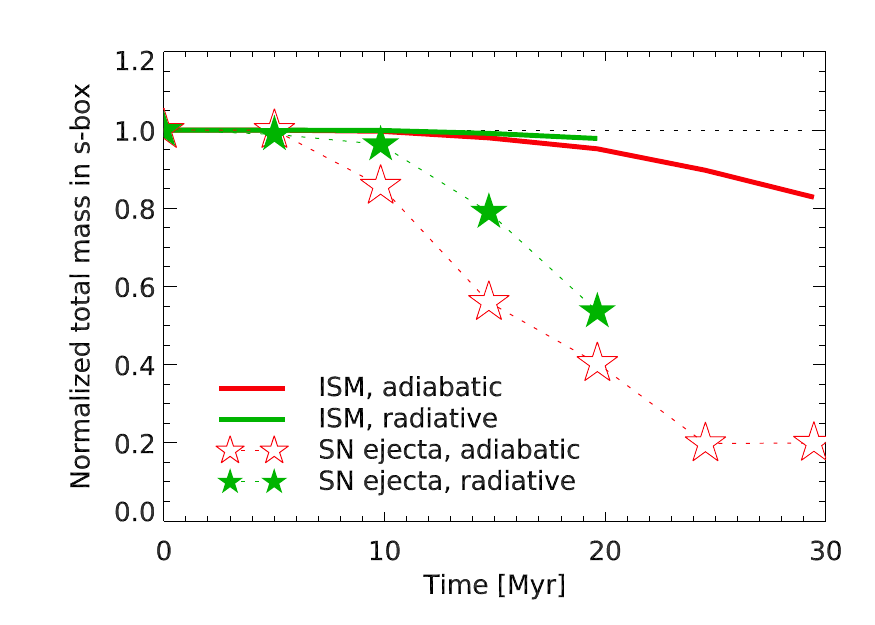}
   \caption{ Cumulative mass (upper panel) and metals (middle panel) ejected by 
     massive stars ($m >$ 8~M$_\odot$) and recycled in the simulation volume, at 
     different times. The (red) empty stars refer to the adiabatic run, the 
     (green) filled stars to the run with radiative cooling. The quantities 
     computed using injection rates from \citet[][]{2014ApJS..212...14L} are 
     shown as dashed (black) lines in all panels, and would be recovered in the 
     case of a closed system. Also shown (lower panel) are the total gas masses 
     (normalized to the initial gas mass of the system; solid lines) and SN 
     ejecta (normalized to the corresponding quantities computed with 
     \citealt[][]{2014ApJS..212...14L} mass return rate at each time; stars 
     joined by dotted lines) in the system at different times, for the 
     adiabatic (red lines and symbols) and for the radiative model (green lines 
     and symbols).}
   \label{fig:mult}
   \end{figure}

\subsection{Hydrodynamic evolution driven by stellar activity}
\label{sec:star}

   At the beginning of the simulation, because of the relatively low densities 
   ($n_{\rm{H}} \simeq$ 7~cm$^{-3}$ at the centre and $\sim$1~cm$^{-3}$ at the 
   Plummer radius) radiative losses are fairly ineffective. Therefore, large 
   cavities filled with tenuous, hot ($T \ge$~10$^8$~K) gas are carved out 
   around the OB associations already during the pre-SN phase (0--3~Myr), both 
   in the adiabatic and cooling models (Figs.~\ref{fig:mapdtz1} and 
   \ref{fig:mapdtz2}, upper panels), without any need for switching off cooling 
   artificially in the radiative simulation. During the SN phase (3--30~Myr), 
   in the presence of radiative cooling multiple SN explosions dig superbubbles 
   smaller than those moulded in the adiabatic limit, filled with hot 
   (T~$>$~10$^7$~K), rarefied gas (cfr., in particular, the middle and lower 
   panels in Figs.~\ref{fig:mapdtz1} and \ref{fig:mapdtz2}, respectively, 
   referring to the $t =$~20~Myr snapshots). These bubbles also take more time 
   to lose their individuality, disrupt, and merge. We note that, in spite of 
   the growth of Rayleigh-Taylor instabilities at the borders, the bubbles 
   preserve ovoid shapes delimited by thin, dense, cold shells, until they do 
   not interact with each other. The largest holes are created in the galaxy's 
   outskirts, where the ISM is less dense and, thus, less resilient and less 
   reluctant to being pulled away. Given the low SN rate, however, the filling 
   factor of the superbubbles is overall small and the OB associations fail to 
   produce a large-scale, coherent action of gas removal. Undoubtedly, the 
   adopted geometry of the system, distribution of OB associations and static 
   implementation of stellar feedback play a role in determining the exact 
   amount of gas that is expelled from the system. We briefly touch upon this 
   in Section~\ref{sec:misc}. The issue of the dependence of the results on 
   variations in number and position of (static) OB associations is discussed 
   in more detail in Appendix~\ref{sec:AppA}.

   Figs.~\ref{fig:mapvr20} and \ref{fig:mapvr30} depict, respectively, the gas 
   radial velocity field at $t = 20$~Myr (for the adiabatic and radiative 
   simulations) and $t = 30$~Myr (only for the adiabatic simulation). Pockets 
   of tenuous, heated gas around individual OB associations expand at 
   supersonic velocity. SN debris are channelled in hot structures, that 
   resemble chimneys and fountains, and can either be entrained in an outflow 
   or `rain back' to the galaxy centre with velocities up to several hundreds 
   of km~s$^{-1}$. The cooler ambient medium is largely unaffected by the SN 
   activity and remains bound in both the radiative and adiabatic regimes; 
   however, in the latter case a fraction of the gas, mostly in the outermost 
   regions, is swept up and steadily moves outwards with velocities that exceed 
   the local escape velocity\footnote{Following \citet[][their 
       equation~1]{2018MNRAS.481.4877N}, the escape velocity is computed by 
     taking into account both the energy required to escape from the `surface' 
     of the halo and that needed to reach the surface from inner regions.} (see 
   insets in Figs.~\ref{fig:mapvr20} and \ref{fig:mapvr30}). 

   The maps in Figs.~\ref{fig:mapvr20}--\ref{fig:mapvr30} capture a limited 
   portion of the simulation volume. A comprehensive view of the total gas mass 
   participating in the outflow is provided in Fig.~\ref{fig:mwvd}, where we 
   show the mass fraction of gas (normalized to the total gas mass in the 
   simulation volume) per velocity (upper panels) and density bin (middle 
   panels), for the adiabatic and radiative runs (left-hand and right-hand 
   panels, respectively) at $t = 20$~Myr. The distribution for all gas is 
   divided into the contribution from a cold/warm ($T <$ 10$^5$~K) and a hot 
   phase ($T \ge$ 10$^5$~K). While most of the outflowing material is in a hot 
   phase with velocities in excess of a few hundreds km~s$^{-1}$, the cold/warm 
   component peaks at small ($|\vel_{\, \rm{r}}| <$ 50~km~s$^{-1}$) velocities. In 
   the radiative case, the hot gas is characterised by a considerably narrower 
   velocity distribution, with the absence of a tail at very high 
   ($|\vel_{\, \rm{r}}| >$ 500~km~s$^{-1}$) velocities. 

   The majority of the gas in the simulation volume does not contribute to the 
   mass carried away by the wind: contrary to conventional wisdom (see 
   Section~\ref{sec:hyvsan}), even in the adiabatic limit we find that after 
   30~Myr the system has lost less than 20 percent of its initial gaseous mass 
   (see Fig.~\ref{fig:mult}, lower panel).

   As already stressed elsewhere \citep[e.g.,][]{2002ApJ...571...40M}, there 
   are two main reasons why the energy deposition, although being higher than 
   the gas binding energy, does not produce a complete blow-away of the gas, 
   even in the adiabatic regime: in real galaxies, the explosion sites are 
   scattered, rather than being all closely packed at the centre. The energy 
   released in the outer, low-density regions tends to escape the galactic 
   potential well without much coupling with the ambient medium, while 
   off-nuclear explosions in the inner regions tend to compress the gas at the 
   centre through inward-propagating shocks (which enhances the radiative 
   losses in the cooling model). In Appendix~\ref{sec:AppA} we show that the 
   system can get rid of a higher fraction of gas if the OB associations are 
   more concentrated (although we do not force them to lie all at the centre of 
   the system, because this would produce a nuclear star cluster, which is not 
   observed in local UFDs). Overall, from our simulations we conclude that in 
   order to sweep the galaxy clean of its gas some external force/event has 
   likely to be considered.

\subsection{Metal enrichment}
\label{sec:met}

   Even if the galaxy does not get rid of its cold gas, most of the mass and 
   energy injected by massive stars in the ISM escape the system.

   The upper and middle panels of Fig.~\ref{fig:mwvd} show that, at $t =$ 
   20~Myr, most of the gas is in the cold/warm component and has relatively 
   large densities ($>$~10$^{-2}$~cm$^{-3}$). The same is not true for the 
   metals. The majority of the metals reside in the hot component 
   (Fig.~\ref{fig:mwvd}, lower panels). In particular, in the adiabatic case 
   their velocity distribution is much broader than the one of the gas, without 
   any substantial dominance of the low-velocity component. This implies that 
   the metals are not mixed with the cold, dense pristine gas. As can be seen 
   from Fig.~\ref{fig:mult}, lower panel, for the adiabatic run at $t =$ 20~Myr 
   only 40\% of massive stars' ejecta is trapped within the simulation volume, 
   and this percentage drops to 20\% at 30~Myr\footnote{For the lower 
     resolution run with more concentrated SN explosions a much higher metal 
     retention if found during most of the simulation (see 
     Fig.~\ref{fig:ejecvg}, Appendix~\ref{sec:AppA}).}. For the run with 
   radiative cooling, 55\% of the stellar ejecta is still present in the 
   simulation box at $t =$ 20~Myr, but this percentage is likely to decrease at 
   later times.

   \begin{figure}
   \centering
   \includegraphics{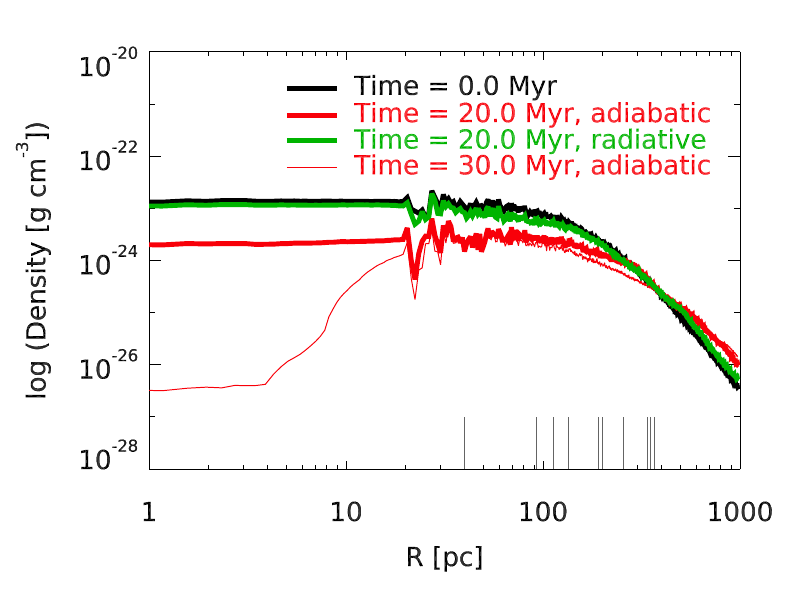}
   \caption{ Initial (black line) and final density profiles, for the adiabatic 
     simulation ($t = 30$~Myr; thin red line) and for the run with radiative 
     cooling ($t = 20$~Myr; thick green line). Also shown is the density 
     profile for the adiabatic simulation at $t = 20$~Myr (thick red line). 
     The vertical lines on the bottom indicate the location of the OB 
     associations.}
   \label{fig:densprof}
   \end{figure}

   \begin{figure}
   \centering
   \includegraphics{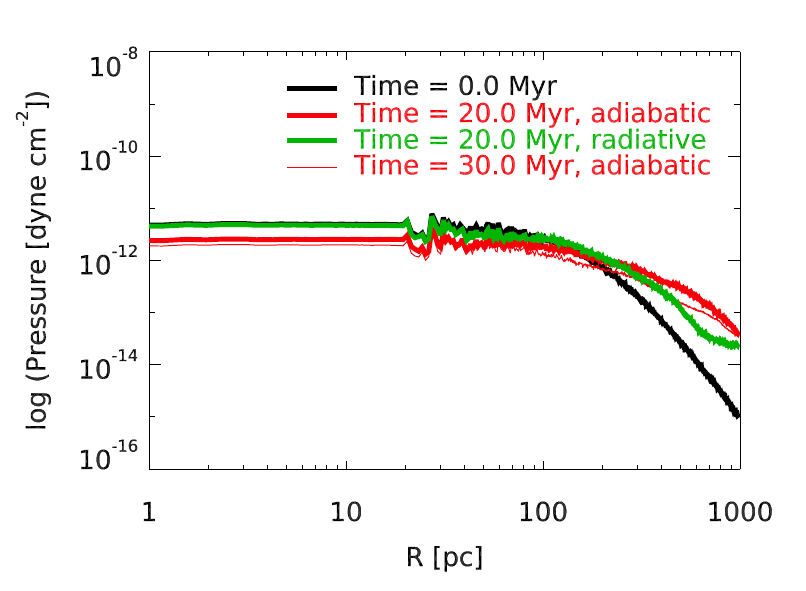}
   \caption{ Same as Fig.~\ref{fig:densprof}, but for the pressure profiles.}
   \label{fig:presprof}
   \end{figure}

   \begin{figure}
   \centering
   \includegraphics{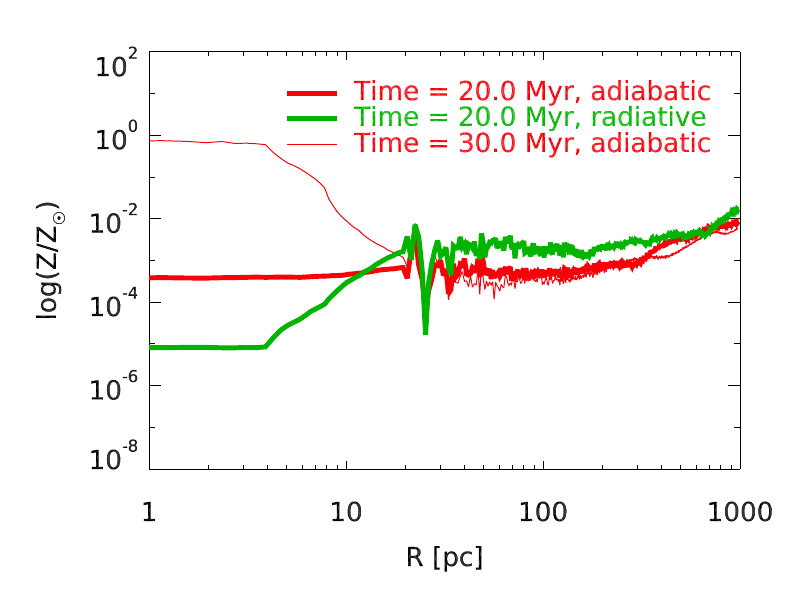}
   \caption{ Metallicity profiles, for the Bo\"otes~I-like UFD simulated in the 
     adiabatic limit (red lines) and with radiative cooling (green line). Thick 
     lines are for $t =$ 20 Myr, the thin one is for $t =$ 30 Myr.}
   \label{fig:Zprof}
   \end{figure}

   \begin{figure}
   \centering
   \includegraphics{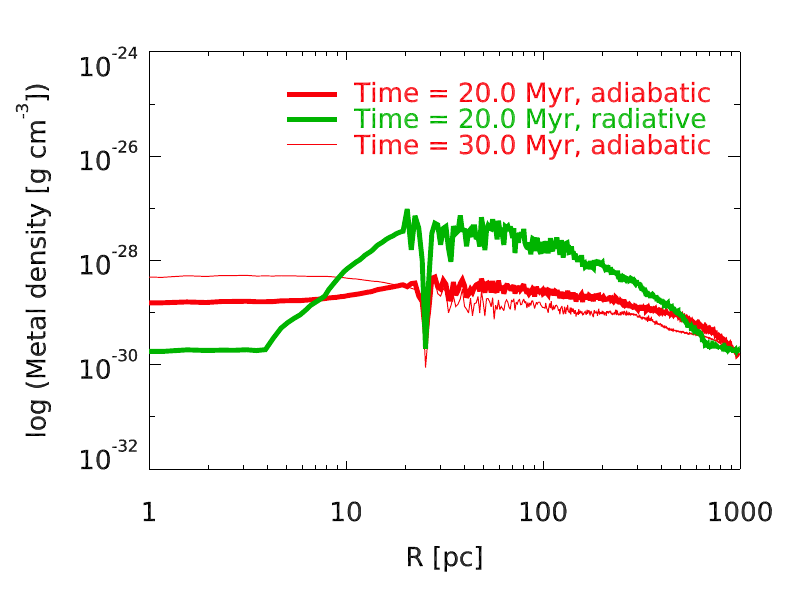}
   \caption{ Same as Fig.~\ref{fig:Zprof}, but the profiles refer to the metal 
     density, $\rho_Z = Z\,\rho_{\rm{gas}}$.}
   \label{fig:densmetprof}
   \end{figure}

   The metal-rich, hot SN ejecta is clearly more easily removed from the system 
   than the cold ambient medium. For the sake of completeness, in the upper and 
   middle panels of Fig.~\ref{fig:mult} we display, respectively, the total SN 
   ejecta and the metals (not-normalised quantities) that are retained in the 
   simulation volume at different times, for the adiabatic (red lines and 
   symbols) and cooling (green lines and symbols) models. These can be directly 
   compared to the corresponding quantities computed using the mass injection 
   rates from \citet[][]{2014ApJS..212...14L}, to be recovered in the case of a 
   closed system (dashed black lines).

\subsection{Radial profiles}
\label{sec:rad}

   The evolution of the gas density profile is shown in 
   Fig.~\ref{fig:densprof}. For the simulation with radiative cooling, the 
   profile at $t = 20$~Myr (green thick curve) is undistinguishable from the 
   initial one (black solid curve), which means that, to a large extent, the 
   ambient medium is unaffected by the mass return from SNe. On the other hand, 
   the density profiles for the galaxy simulated in the adiabatic regime at 
   $t = 20$~Myr (red thick curve) and 30~Myr (red thin curve) deviate 
   considerably from the initial one: in particular, the final profile displays 
   densities three orders of magnitude lower at the centre (i.e., within the 
   innermost 10~pc) and slightly higher beyond 500~pc. The pressure profiles 
   clearly show that the hot pressurized SN ejecta is expanding to cross the 
   simulation box boundaries at late times (Fig.~\ref{fig:presprof}).

   Owing to the larger filling factor of the superbubbles (cf. 
   Figs.~\ref{fig:mapdtz1} and \ref{fig:mapdtz2}) and to the more efficient 
   loss of metals (Fig.~\ref{fig:mult}, bottom panels), at $t =$ 20~Myr the 
   metallicity profile predicted in the adiabatic limit is flatter than that 
   predited with the cooling model (Fig.~\ref{fig:Zprof}), and less metals are 
   found beyond the innermost 10~pc (Fig.~\ref{fig:densmetprof}). In the 
   radiative regime, the metals tend to remain close to the regions where they 
   are injected by dying stars, which explains the rise from the inner regions 
   (we remind the reader that for the high-resolution runs discussed in this 
   section the OB associations are randomly distributed following a Plummer's 
   mass profile with $a \simeq$ 200~pc, see Section~\ref{sec:setup}). The 
   decrease of the metal density profile at large radii is due to the leakage 
   of metals from the simulation box. At later times ($t =$ 30~Myr), a 
   low-density, metal-rich blob occupies the central regions of the adiabatic 
   model, that is when and where the filling factor of the bubbles approaches 
   unity.

\section{Discussion}
\label{sec:disc}

   In general, the smallest MW satellites nowadays appear gas-free. 
   Understanding the physical processes which caused the removal of their gas 
   is therefore necessary to explain their present-day properties. 

   In pure chemical evolution models, it is often taken for granted that the 
   injection of energy by SNe regulates the onset of galactic-scale outflows. 
   Consequently, the timescales for gas removal, star formation quenching and, 
   ultimately, chemical enrichment are dictated by uncertain prescriptions on 
   stellar feedback. 

   By means of our 3D hydrodynamical simulations, we have shown that in 
   realistic conditions, it appears difficult to explain the removal of the 
   bulk of the cold gas by means of stellar feedback alone. This is at variance 
   with results purely based on analytical calculations, that implicitly assume 
   an idealised configuration with all SNe at the centre and full energy 
   coupling with the gas. We pick up on this point further in the coming 
   paragraphs.

\subsection{Supernova-driven outflows: analytic model expectations versus 
  simulations}
\label{sec:hyvsan}

   In this section, we contrast the expectations about SN-driven outflows from 
   an analytic model for B\"ootes~I with the results of our hydrodynamical 
   simulations performed in the adiabatic regime. 

   The analytic model rests on the original set-up described in 
   Section~\ref{sec:setup}. The binding energy of the gas is given by
   \begin{equation}
   \label{eq:be}
     E_{\rm{g}} = \int_{0}^{\infty} \Phi_{\rm{DM}}(r) \, 
                4 \pi r^2 \rho_{\rm{gas}}(r) \, {\rm d} r.
     \end{equation}
   We assume that the density distribution of the dark matter halo is 
   represented by Burkert's (\citeyear[][]{1995ApJ...447L..25B}) profile, and 
   calculate the potential by integrating Poisson's equation as
   \begin{equation}
     \begin{aligned}
     \Phi_{\rm{DM}}(r) = & -1.96\, \frac{G M_0}{r_0} \, \bigg\{ \pi - 2 \, 
                       \bigg( 1 + \frac{r_0}{r} \bigg) \, 
                       {\rm arctan}\frac{r}{r_0} + \\
                       & + 2 \, \bigg( 1 + \frac{r_0}{r} \bigg) \, 
                       {\rm ln}\bigg( 1 + \frac{r}{r_0} \bigg) - \\
                       & - \bigg( 1 - \frac{r_0}{r} \bigg) \, 
                       {\rm ln}\bigg[ 1 + \bigg( \frac{r}{r_0} \bigg)^2 
                       \bigg] \bigg\},
     \end{aligned}
     \end{equation}
   where $G$ is the gravitational constant, $r_0$ is the core radius of the 
   dark matter distribution, and $M_0$ is the total dark mass inside $r_0$. 
   The observed scaling relation derived by \citet[][]{1995ApJ...447L..25B} 
   yields: 
   \begin{equation}
     \label{eq:r0}
     r_0 = 3.07 \, \bigg( \frac{M_0}{10^9 \, {\rm M}_\odot} \bigg)^{3/7} \, 
     {\rm kpc}
     \end{equation}
   and
   \begin{equation}
     \label{eq:M0}
     M_0 = \frac{M_{\rm DM}}{5.8} \, {\rm M}_\odot,
     \end{equation}
   where $M_{\rm DM} = 3.5 \times 10^7$~M$_\odot$ \citep{2010MNRAS.406.1220W}. 
   Beyond the dark matter cut-off radius, $R_{\rm c}$, the potential simply 
   reads:
   \begin{equation}
     \Phi_{\rm{DM}}(r) = - \frac{G M_{\rm DM}}{r}.
     \end{equation}
   For the gas, we use a Plummer's (\citeyear[][]{1911MNRAS..71..460P}) density 
   distribution,
   \begin{equation}
     \rho_{\rm{gas}}(r) = \frac{3 \, M_{\rm gas}}{4 \, \pi \, a^3} \, \bigg( 
     1 \, + \, \frac{r^2}{a^2} \bigg)^{-5/2},
     \end{equation}
   with a characteristic radius $a \simeq$ 200~pc and 
   $M_{\rm gas} = f_{\, \rm b} \, M_{\rm DM} = 6 \times 10^6$~M$_\odot$, where 
   $f_{\, \rm b}$ is the cosmic baryon fraction \citep{2009ApJS..180..330K}. By 
   integrating Eq.(\ref{eq:be}) with these assumptions, it turns out that
   \begin{equation}
     E_{\rm{g}} \sim 50 \times 10^{51} \, {\rm erg}.
     \end{equation}
   By assuming an average SN energy of $10^{51}$~erg, it is concluded that, in 
   adiabatic conditions, $\sim$50 SNe suffice to completely remove the gas from 
   the galaxy. Our adiabatic simulations, however, tell us a different story. 
   From an inspection of Figs.~\ref{fig:mapdtz1}, \ref{fig:mapvr20} and 
   \ref{fig:mapvr30}, in fact, it is clear that the energy release from SNe 
   clustered in associations is highly spatially inhomogeneous. The SN ejecta 
   are channelled outward through different funnels. If the OB associations are 
   located at large radii, the rarefied gas crosses the boundaries of the 
   computational box and is lost from the system. If, instead, the hot gas 
   originates from the innermost OB associations, it may receive not enough 
   energy to become unbound and, eventually, it may fall to the centre (see 
   Fig.~\ref{fig:mapvr30}, bottom right). The more concentrated the OB 
   associations, the higher the gas fraction that is lost from the system (see 
   Appendix~\ref{sec:AppA}), but we never predict a complete blow-away of the 
   gas.

   In summary, under our specific premises on the initial distributions of gas 
   and stars we show that, though the number of SNe that explode in the 
   simulated galaxy is one order of magnitude larger than required to blow away 
   the entire gas reservoir according to the analytical estimate, a global wind 
   is unlikely to develop \citep[see also][]{2002ApJ...571...40M}. Although the 
   outflow may entrain a non-negligible fraction of the ambient ISM, the galaxy 
   is not swept clean of its gas in our simulations.
   
\subsection{The role of stellar feedback in the evolution of dwarf galaxies}

   In principle, our results may pose a problem for the evolution of dwarf 
   galaxies on a larger perspective. One crucial aspect concerns the capability 
   of the stellar feedback of blowing away the gas and regulating star 
   formation in small systems. In semi-analytic and pure chemical evolution 
   models, a simple criterion for star formation regulation by stellar feedback 
   is commonly assumed, i.e. that the ISM might be ejected from a galaxy when 
   the cumulative energy injected by all SNe ever exploded exceeds the binding 
   energy of the remaining gas \citep[e.g.][]{1974MNRAS.169..229L,
     1994A&ARv...6...67F}. Our results strengthen previous findings in the 
   literature \citep[e.g.][]{2006MNRAS.371..643M,2015ApJ...805..109C} that such 
   a criterion could be too simplistic, and that perhaps the role of stellar 
   feedback in the regulation of star formation in dwarf galaxies needs to be 
   reconsidered. 

   In the past, it has been also pointed out that the compression due to a 
   large accumulation of energy from nearby OB associations might lead to very 
   high gas densities, with very rapid cooling and dissipation of such energy 
   and, consequently, inefficient effects of feedback 
   \citep[e.g.][]{1986ApJ...305..669V}. Our simulations have shown clearly that 
   radiative cooling does not play a major role in the failure of the removal 
   of the gas by stellar feedback. In fact, with our assumptions, even in 
   adiabatic conditions, i.e. in the complete absence of radiative cooling, 
   stellar winds and SNe are unable to power a massive wind capable of removing 
   the bulk of the cold gas from a Bo\"otes~I-like galaxy. Indeed, the energy 
   injected in the ISM tends to escape the system along `privileged paths', 
   rather than be uniformly spread and used to heat the gas all over the 
   simulation volume.

   One major future development of our study will be a direct assessment of the 
   effects of stellar feedback in star-forming UFDs, simulated at very high 
   spatial resolution. A relevant previous attempt to simulate isolated 
   gas-rich dwarf galaxies taking into account their star formation history, 
   which also includes a halo in the UFD domain, has been done by 
   \citet{2016MNRAS.459.2573R}, which resolve the interactions of the bubbles 
   created by individual SN explosions \citep[for high-resolution cosmological 
     simulations, see][]{2018arXiv181202749W}. The authors of that work state 
   that their lowest mass halo forms virtually no stars, so they refrain from 
   further analysis, but this could perhaps imply that it clears enough gas to 
   quench the star formation in the system. Detailed high-resolution, 
   three-dimensional simulations of UFDs including both SN feedback and star 
   formation history are necessary to investigate further these crucial aspects 
   of dwarf galaxy evolution at the lowest mass end, as well as to gain a 
   broader understanding of such a disparate class of objects.

\subsection{On galactic winds in dwarf galaxies and massive star clusters}
\label{sec:misc}

   For a long time, energetic feedback from stellar winds and SNe have known to 
   be able to drive vast superbubbles. A direct observational confirmation of 
   the presence of superbubbles in a local system was possible thanks to 
   spectacular, high-spatial resolution Chandra X-ray images of the 30~Doradus 
   star-forming complex in the Large Magellanic Cloud 
   \citep{2006AJ....131.2140T}. This study provided a clear view of the complex 
   network generated by interacting stellar winds and SNe, working together to 
   create large cavities filled with hot X-ray emitting plasma. On the 
   theoretical side, the true impact of such powerful events on the evolution 
   of the smallest stellar systems, such as dwarf galaxies and stellar 
   clusters, is still an open question.

   Due to their shallow potential wells and, in some cases, low surface 
   brightness, which implies low stellar and gas densities, dwarf galaxies 
   should in principle be highly vulnerable to the effects of stellar winds and 
   SN feedback. Several previous studies already pointed out that the smallest 
   systems (i.e those with total mass smaller than about $10^7$~M$_{\odot}$) 
   should more easily have most of their natal gas removed by SN-driven 
   large-scale outflows \citep[e.g.][]{1999ApJ...513..142M,2013A&A...551A..41R}.

   A recent numerical study of metal-rich winds in dwarf galaxies is the one of 
   \citet{2017ApJ...835..136R}, based on 3D $N$-body/smoothed particle 
   hydrodynamics. The study focuses the attention on important parameters such 
   as the galaxy concentration index, the gas fraction and the shape of the 
   mass distribution, as well as the position of the starburst inside the 
   system. These authors find that an off-center starburst in dwarf galaxies is 
   the most effective mechanism to produce a significant loss of metals. In 
   their comprehensive survey of models characterised by different parameters, 
   they find that most winds produced by starbursts in dwarf galaxies have a 
   high metal content, but that in general the winds are not well-mixed, in 
   that the metals ejected by the massive stars are unable to mix efficiently 
   with the gas of the galaxy.

   In other studies, the shape of the galaxy seems to be key in gas expulsion. 
   For instance, \citet{2001ApJ...552...91S} concluded that the ejection of 
   enriched gas from a flat galaxy is facilitated with respect to a spherical 
   galaxy \citep[see also][]{2013A&A...551A..41R}. From such studies, it was 
   concluded that the fate of the pristine gas is generally more dependent on 
   the total mass (including a dark matter halo), and that smaller galaxies 
   develop larger outflows, and in these systems the fractions of the ejected 
   gas tends to be larger.

   It is interesting to note that also in massive stellar clusters the 
   capability of superbubbles to drive outflows is a controversial issue, in 
   particular in systems with baryonic mass comparable to that of Bo\"otes~I. 
   Massive clusters with total mass $10^7$~M$_{\odot}$ are much more compact 
   than a dwarf galaxy, are not dark matter dominated and have generally a 
   factor of ten more massive stars than Bo\"otes~I. In principle, such stars 
   could contribute simultaneously to the creation of a large-scale outflow; in 
   the work of \citet{2008MNRAS.384.1231B}, these conditions were sufficient to 
   unbind the gas only in systems with mass smaller than $10^7$~M$_{\odot}$ 
   \citep[see also][]{1986ApJ...304..283D,2016A&A...587A..53K}.
   
   On the other hand, in \citet{2015ApJ...814L..14C} the evolution of a massive 
   cluster with total mass $10^7$~M$_{\odot}$ was studied by means of 
   three-dimensional hydro-simulations very similar to the ones carried out in 
   this work. A distribution of OB associations scattered in the cluster was 
   taken into account, and with the thermal modelling of the ISM considered in 
   that paper, which included both heating by stellar winds and SNe and 
   radiative cooling, the conclusion was that the entire initial gas content is 
   blown away already after 15~Myr of evolution. Other studies are needed to 
   investigate what is the fate of the gas in such system if different 
   implementations of stellar feedback are taken into account, to test, e.g., 
   the effects of the injection of momentum or of a combination of thermal 
   energy and momentum. In the study of \citet{2015ApJ...814L..14C}, what is 
   key for the ejection of the gas is the coherence of concentrated 
   associations, which causes the hot gas to cover a significant fraction of 
   the volume in a short time (a few Myr). This seems to be the condition to 
   drive a steady wind, as also other studies have shown 
   \citep[e.g.][]{2017MNRAS.465.1720Y}, and in which a sufficiently large 
   ($10^4$, or more) number of SNe acting simultaneously seems to be required. 
   In the light of these findings, and also following the insight provided 
     by the lower resolution simulations presented in Appendix~\ref{sec:AppA} 
     of this paper, it seems that one major cause of the failure of stellar 
   feeback in driving a steady mass outflow might be the relative isolation of 
   the OB associations.

   Finally, a few considerations are in order concerning our static 
   implementation of stellar feedback. In principle, these conditions should 
   maximize the effects of stellar feedback with respect to particle-like (i.e. 
   free to move) energy sources. In fact, in a medium already heated and 
   diluted by previous activity of stellar winds as the one of our simulation, 
   the action of the SN feedback should be enhanced, as the energy released by 
   both sources can accumulate around their fixed positions. If OB associations 
   were free to move and if their dynamics were followed in detail, it is 
   likely that their effect on the system would be less pronounced 
   \citep{2015A&A...579A...9V}.

\subsection{Other numerical issues}

   In a simulated high-density gas, the energy deposited by massive stars in 
   the pre-SN and SN phases can be radiated away very quickly, which renders 
   the stellar feedback highly inefficient \citep{1992ApJ...391..502K}. This 
   implies that other assumptions are needed in order to have an appreciable 
   effect of stellar winds and SNe on the ISM \citep[see, 
     e.g.,][]{2000ApJ...545..728T,2013ApJ...770...25A} by, e.g., injecting 
   momentum in fully radiative conditions, or by switching off cooling in an 
   appropriate and hopefully realistic fashion (such as, for instance, in some 
   `spheres of influence' around OB associations). This problem seems to be 
   related to our poor understanding of the physical processes associated with 
   stellar feedback which, to be described realistically, would require several 
   ingredients particularly complex to implement, including magnetic fields, 
   turbulence and other non-thermal processes such as, e.g., cosmic rays, as 
   well as stellar radiation \citep[see][]{2013MNRAS.429.3068T}.

   In this work, we have shown that tenuous, hot bubbles are created also in 
   our radiative simulation; this confirms that at high resolution, the effects 
   of stellar feedback tend to be insensitive to the details of the subgrid 
   physics \citep{2016MNRAS.459.2573R}. Our results strengthen and confirm 
   recent findings of other authors who have shown that, if the `cooling 
   radius' of the interstellar bubbles is resolved well enough, the momentum 
   which accompanies the fast ejecta of the OB association is correctly 
   recovered before the deposited energy is radiated away 
   \citep{2015ApJ...802...99K,2015MNRAS.450..504M,2015ApJ...809...69S}.

   Another remark is in order, which concerns the well-known numerical 
   difficulties that are typical of studies like the present one. In fact, the 
   fast fluid injected by stellar winds and SNe, with typical velocities of a 
   few $10^3$~km~s$^{-1}$, is particularly difficult to treat computationally. 
   This occurs mostly because, in order to satisfy the Courant-Friedrichs-Lewy 
   condition, very small timesteps are generally required (for our 
   high-resolution simulations, ${\rm d}t$ is of the order of 100--400~yr). 
   Some authors have chosen to overcome this difficulty by artificially 
   decreasing the wind velocity \citep[e.g.][]{2019MNRAS.482.1304E}. In our 
   simulations, the wind velocity has not been altered, and we have been able 
   to complete our high-resolution adiabatic run up to 30 Myr. However, we have 
   been unable to follow the entire evolution of our galaxy in the 
   high-resolution radiative simulation, because of too highly demanding 
   computational constraints. At sub-parsec resolution, in fact, the radiative 
   simulation is significantly slowed down: a considerable portion of the 
   domain is refined to the maximum level, hence, at each timestep the implicit 
   algorithm for radiative cooling used within \textsc{ramses} has to run an 
   extremely large amount of times (of the order of 200 million times, 
   corresponding to the number of grids with the highest refinement level). 
   This, however, was not an impediment, since our aim was to study the effects 
   of stellar feeback in an UFD and, in a conservative fashion, we have been 
   able to show that, even in a maximal case of feedback efficiency (i.e. 
   without radiative losses), it is pretty hard to remove the bulk of the gas 
   from the system for a realistic distribution of OB associations (see also 
   the results of our lower resolution numerical experiments in 
   Appendix~\ref{sec:AppA}). In fact, a scattered distribution of the OB 
   associations makes stellar feedback less efficient than expected from simple 
   energetic arguments: cold gas may pile up in some regions, and it is 
   relatively easy to carve tunnels and chimneys along which the massive star 
   energy is vented out of the system, rather than being used to accelerate the 
   ambient gas \citep[see also][]{2002ApJ...571...40M}.

\section{Conclusions}
\label{sec:conc}

   This paper is the first in a series aimed at studying the evolution of the 
   interstellar medium in ultrafaint dwarf galaxies. In particular, our aim is 
   to understand which physical process ultimately caused the removal of gas 
   from a system resembling Bo\"otes~I.
 
   By means of idealized three-dimensional grid-based numerical simulations we 
   have studied the effects of an internal process, i.e. the stellar feedback, 
   taking into account both stellar winds and SN explosions, on the early 
   evolution of Bo\"otes~I. The fast fluid injected by stellar winds and SNe, 
   with typical velocities of a few $10^3$~km~s$^{-1}$, particularly difficult 
   to treat computationally, is fully taken into account. 

   We assumed an instantaneously born stellar population, and that massive 
   stars are grouped in OB associations scattered across the computational 
   volume. Each association was allowed to inject mass and energy in its 
   surroundings at a constant pace for an uninterrupted period of 30~Myr 
   (roughly corresponding to the lifetime of a 8~M$_\odot$ star).

   We have run both adiabatic simultations, in which radiative cooling was 
   switched off, and a radiative simulation. We have run high-resolution and 
   lower resolution simulations. The results of the high-resolution simulations 
   are presented in Section~\ref{sec:res}, while a suite of lower resolution 
   simulations is discussed in Appendix~\ref{sec:AppA}. Our findings can be 
   summarised as follows.
   \begin{itemize}
   \item In the adiabatic case, the effects of stellar feedback are to be 
     regarded as maximal as, in principle, the energy injected by OB 
     associations is entirely used  to heat the ISM. Nevertheless, once the OB 
     associations are randomly distributed over the simulation volume 
     (following the density profile of the gas, or a more concentrated one), 
     the results of the simulations fall short of the expectations from simple 
     energetic arguments. In these conditions, in fact, the energy injected by 
     stellar winds and SNe is not fully coupled to the cold gas present in the 
     system. After 30~Myr, the system has lost less than 20--30 percent of its 
     initial gaseous mass in our simulations, the exact figure depending on the 
     location of the OB associations more than on their number. However, in 
     this time interval the radial distribution of gas is subject to 
     substantial evolution. At the end of the high-resolution simulation, the 
     density has decreased by three orders of magnitude at the centre, and 
     increased in the outskirts. At the same time, the hot, pressurized gas 
     expands to cross the simulation box boundaries. Due to computational 
     reasons, the radiative high-resolution run has been followed for 20~Myr 
     only. In this case, the cold, initial gas is even less affected by stellar 
     feedback. 
   \item In both the adiabatic and radiative case, a substantial amount of the 
     hot ejecta provided by OB associations is lost from the system. In the 
     adiabatic case, at 20~Myr (30~Myr) from 30 to 60 per cent (from 70 to 80 
     per cent) of the ejecta has left the system, the exact amount depending on 
     the specific location of the OB associations. As for the radiative 
     simulation, at 20~Myr nearly 50 per cent of the ejecta has been expelled. 
     Consistently with previous results from the literature, at the end of the 
     simulation the hot, metal-enriched tenuous gas driven by OB associations 
     is not well-mixed with the cold, pristine gas.
   \item Interstellar bubbles are rapidly and efficiently created in both the 
     adiabatic and radiative simulations. This occurs thanks to our high 
     resolution, which renders the effects of stellar feedback insensitive to 
     the details of the subgrid physics \citep{2016MNRAS.459.2573R}. This 
     confirms the results from other recent studies, which have shown that if 
     the `cooling radius' of interstellar bubbles is resolved well enough, 
     momentum happens to be correctly deposited in the medium 
     \citep[e.g.][]{2015ApJ...802...99K}.
   \end{itemize} 
   Finally,  we caution the reader that these results have to be interpreted 
   with caution, always bearing in mind the approximations introduced in the 
   underlying model.

   If OB associations fail to cause the removal of the bulk of the cold gas, 
   environmental processes need to be invoked for the model UFD to evolve to a 
   gas-free system, a conclusion already drawn for more massive dwarf 
   spheroidal galaxies \citep[e.g.][]{2006MNRAS.371..643M,2015ApJ...805..109C}. 
   Together with a more realistic modeling of star formation, these external 
   processes will be addressed in a forthcoming paper. As we already stressed 
   in Section~\ref{sec:setup}, in the framework of the IGIMF theory a prolonged 
   star formation in UFDs should result in fewer SN explosions, and we plan to 
   implement this theory in our simulations.

\begin{acknowledgements}
   We acknowledge the CINECA awards under the ISCRA initiative and under the 
   MoU INAF-CINECA for the availability of high performance computing resources 
   and support. We acknowledge the computing centre of INAF, Osservatorio 
   Astronomico di Catania, under the coordination of the CHIPP project, for the 
   availability of computing resources and support. CGF would like to 
   acknowledge  the Viper HPC facility at the University of Hull. DR and FC are 
   grateful for financial support from INAF PRIN-SKA \emph{``Empowering SKA as 
     a Probe of galaxy Evolution with \ion{H}{I} (ESKAPE-HI)''} program 
   1.05.01.88.04 (PI L.~K.~Hunt). This work benefited also from the 
   International Space Science Institute (ISSI) in Bern, CH, thanks to the 
   funding of the team \emph{``The Formation and Evolution of the Galactic 
     Halo''} (PI D.~Romano). Last but not least, the authors are indebted to 
   the anonymous referee for an extremely thorough report, which helped to 
   significantly improve the manuscript, and to Pavel Kroupa, who commented on 
   an earlier version of this paper.
\end{acknowledgements}

\bibliographystyle{aa} 
\bibliography{/Users/donatella/Papers/pap-3d-boo/R19_3d_boo_BIB}

\begin{thebibliography}{83}
\expandafter\ifx\csname natexlab\endcsname\relax\def\natexlab#1{#1}\fi

\bibitem[{{Abel} {et~al.}(2000){Abel}, {Bryan}, \&
  {Norman}}]{2000ApJ...540...39A}
{Abel}, T., {Bryan}, G.~L., \& {Norman}, M.~L. 2000, \apj, 540, 39

\bibitem[{{Agertz} {et~al.}(2013){Agertz}, {Kravtsov}, {Leitner}, \&
  {Gnedin}}]{2013ApJ...770...25A}
{Agertz}, O., {Kravtsov}, A.~V., {Leitner}, S.~N., \& {Gnedin}, N.~Y. 2013,
  \apj, 770, 25

\bibitem[{{Aihara} {et~al.}(2018){Aihara}, {Arimoto}, {Armstrong}, {Arnouts},
  {Bahcall}, {Bickerton}, {Bosch}, {Bundy}, {Capak}, {Chan}, {Chiba}, {Coupon},
  {Egami}, {Enoki}, {Finet}, {Fujimori}, {Fujimoto}, {Furusawa}, {Furusawa},
  {Goto}, {Goulding}, {Greco}, {Greene}, {Gunn}, {Hamana}, {Harikane},
  {Hashimoto}, {Hattori}, {Hayashi}, {Hayashi}, {He{\l}miniak}, {Higuchi},
  {Hikage}, {Ho}, {Hsieh}, {Huang}, {Huang}, {Ikeda}, {Imanishi}, {Inoue},
  {Iwasawa}, {Iwata}, {Jaelani}, {Jian}, {Kamata}, {Karoji}, {Kashikawa},
  {Katayama}, {Kawanomoto}, {Kayo}, {Koda}, {Koike}, {Kojima}, {Komiyama},
  {Konno}, {Koshida}, {Koyama}, {Kusakabe}, {Leauthaud}, {Lee}, {Lin}, {Lin},
  {Lupton}, {Mandelbaum}, {Matsuoka}, {Medezinski}, {Mineo}, {Miyama},
  {Miyatake}, {Miyazaki}, {Momose}, {More}, {More}, {Moritani}, {Moriya},
  {Morokuma}, {Mukae}, {Murata}, {Murayama}, {Nagao}, {Nakata}, {Niida},
  {Niikura}, {Nishizawa}, {Obuchi}, {Oguri}, {Oishi}, {Okabe}, {Okamoto},
  {Okura}, {Ono}, {Onodera}, {Onoue}, {Osato}, {Ouchi}, {Price}, {Pyo}, {Sako},
  {Sawicki}, {Shibuya}, {Shimasaku}, {Shimono}, {Shirasaki}, {Silverman},
  {Simet}, {Speagle}, {Spergel}, {Strauss}, {Sugahara}, {Sugiyama}, {Suto},
  {Suyu}, {Suzuki}, {Tait}, {Takada}, {Takata}, {Tamura}, {Tanaka}, {Tanaka},
  {Tanaka}, {Tanaka}, {Terai}, {Terashima}, {Toba}, {Tominaga}, {Toshikawa},
  {Turner}, {Uchida}, {Uchiyama}, {Umetsu}, {Uraguchi}, {Urata}, {Usuda},
  {Utsumi}, {Wang}, {Wang}, {Wong}, {Yabe}, {Yamada}, {Yamanoi}, {Yasuda},
  {Yeh}, {Yonehara}, \& {Yuma}}]{2018PASJ...70S...4A}
{Aihara}, H., {Arimoto}, N., {Armstrong}, R., {et~al.} 2018, \pasj, 70, S4

\bibitem[{{Baumgardt} {et~al.}(2008){Baumgardt}, {Kroupa}, \&
  {Parmentier}}]{2008MNRAS.384.1231B}
{Baumgardt}, H., {Kroupa}, P., \& {Parmentier}, G. 2008, \mnras, 384, 1231

\bibitem[{{Bechtol} {et~al.}(2015){Bechtol}, {Drlica-Wagner}, {Balbinot},
  {Pieres}, {Simon}, {Yanny}, {Santiago}, {Wechsler}, {Frieman}, {Walker},
  {Williams}, {Rozo}, {Rykoff}, {Queiroz}, {Luque}, {Benoit-L{\'e}vy},
  {Tucker}, {Sevilla}, {Gruendl}, {da Costa}, {Fausti Neto}, {Maia}, {Abbott},
  {Allam}, {Armstrong}, {Bauer}, {Bernstein}, {Bernstein}, {Bertin}, {Brooks},
  {Buckley-Geer}, {Burke}, {Carnero Rosell}, {Castander}, {Covarrubias},
  {D'Andrea}, {DePoy}, {Desai}, {Diehl}, {Eifler}, {Estrada}, {Evrard},
  {Fernandez}, {Finley}, {Flaugher}, {Gaztanaga}, {Gerdes}, {Girardi},
  {Gladders}, {Gruen}, {Gutierrez}, {Hao}, {Honscheid}, {Jain}, {James},
  {Kent}, {Kron}, {Kuehn}, {Kuropatkin}, {Lahav}, {Li}, {Lin}, {Makler},
  {March}, {Marshall}, {Martini}, {Merritt}, {Miller}, {Miquel}, {Mohr},
  {Neilsen}, {Nichol}, {Nord}, {Ogando}, {Peoples}, {Petravick}, {Plazas},
  {Romer}, {Roodman}, {Sako}, {Sanchez}, {Scarpine}, {Schubnell}, {Smith},
  {Soares-Santos}, {Sobreira}, {Suchyta}, {Swanson}, {Tarle}, {Thaler},
  {Thomas}, {Wester}, {Zuntz}, \& {DES Collaboration}}]{2015ApJ...807...50B}
{Bechtol}, K., {Drlica-Wagner}, A., {Balbinot}, E., {et~al.} 2015, \apj, 807,
  50

\bibitem[{{Bellazzini} {et~al.}(2018){Bellazzini}, {Armillotta}, {Perina},
  {Magrini}, {Cresci}, {Beccari}, {Battaglia}, {Fraternali}, {de Zeeuw},
  {Martin}, {Calura}, {Ibata}, {Coccato}, {Testa}, \&
  {Correnti}}]{2018MNRAS.476.4565B}
{Bellazzini}, M., {Armillotta}, L., {Perina}, S., {et~al.} 2018, \mnras, 476,
  4565

\bibitem[{{Belokurov} {et~al.}(2014){Belokurov}, {Irwin}, {Koposov}, {Evans},
  {Gonzalez-Solares}, {Metcalfe}, \& {Shanks}}]{2014MNRAS.441.2124B}
{Belokurov}, V., {Irwin}, M.~J., {Koposov}, S.~E., {et~al.} 2014, \mnras, 441,
  2124

\bibitem[{{Belokurov} {et~al.}(2007){Belokurov}, {Zucker}, {Evans}, {Kleyna},
  {Koposov}, {Hodgkin}, {Irwin}, {Gilmore}, {Wilkinson}, {Fellhauer},
  {Bramich}, {Hewett}, {Vidrih}, {De Jong}, {Smith}, {Rix}, {Bell}, {Wyse},
  {Newberg}, {Mayeur}, {Yanny}, {Rockosi}, {Gnedin}, {Schneider}, {Beers},
  {Barentine}, {Brewington}, {Brinkmann}, {Harvanek}, {Kleinman}, {Krzesinski},
  {Long}, {Nitta}, \& {Snedden}}]{2007ApJ...654..897B}
{Belokurov}, V., {Zucker}, D.~B., {Evans}, N.~W., {et~al.} 2007, \apj, 654, 897

\bibitem[{{Belokurov} {et~al.}(2006){Belokurov}, {Zucker}, {Evans},
  {Wilkinson}, {Irwin}, {Hodgkin}, {Bramich}, {Irwin}, {Gilmore}, {Willman},
  {Vidrih}, {Newberg}, {Wyse}, {Fellhauer}, {Hewett}, {Cole}, {Bell}, {Beers},
  {Rockosi}, {Yanny}, {Grebel}, {Schneider}, {Lupton}, {Barentine},
  {Brewington}, {Brinkmann}, {Harvanek}, {Kleinman}, {Krzesinski}, {Long},
  {Nitta}, {Smith}, \& {Snedden}}]{2006ApJ...647L.111B}
{Belokurov}, V., {Zucker}, D.~B., {Evans}, N.~W., {et~al.} 2006, \apjl, 647,
  L111

\bibitem[{{Bland-Hawthorn} {et~al.}(2015){Bland-Hawthorn}, {Sutherland}, \&
  {Webster}}]{2015ApJ...807..154B}
{Bland-Hawthorn}, J., {Sutherland}, R., \& {Webster}, D. 2015, \apj, 807, 154

\bibitem[{{Brown} {et~al.}(2014){Brown}, {Tumlinson}, \&
  {Geha}}]{2014ApJ...796...91B}
{Brown}, T.~M., {Tumlinson}, J., \& {Geha}, M. e.~a. 2014, \apj, 796, 91

\bibitem[{{Burkert}(1995)}]{1995ApJ...447L..25B}
{Burkert}, A. 1995, \apjl, 447, L25

\bibitem[{{Calura} {et~al.}(2015){Calura}, {Few}, {Romano}, \&
  {D'Ercole}}]{2015ApJ...814L..14C}
{Calura}, F., {Few}, C.~G., {Romano}, D., \& {D'Ercole}, A. 2015, \apjl, 814,
  L14

\bibitem[{{Caproni} {et~al.}(2015){Caproni}, {Lanfranchi}, {da Silva}, \&
  {Falceta-Gon{\c c}alves}}]{2015ApJ...805..109C}
{Caproni}, A., {Lanfranchi}, G.~A., {da Silva}, A.~L., \& {Falceta-Gon{\c
  c}alves}, D. 2015, \apj, 805, 109

\bibitem[{{Chambers} {et~al.}(2016){Chambers}, {Magnier}, {Metcalfe},
  {Flewelling}, {Huber}, {Waters}, {Denneau}, {Draper}, {Farrow}, {Finkbeiner},
  {Holmberg}, {Koppenhoefer}, {Price}, {Saglia}, {Schlafly}, {Smartt},
  {Sweeney}, {Wainscoat}, {Burgett}, {Grav}, {Heasley}, {Hodapp}, {Jedicke},
  {Kaiser}, {Kudritzki}, {Luppino}, {Lupton}, {Monet}, {Morgan}, {Onaka},
  {Stubbs}, {Tonry}, {Banados}, {Bell}, {Bender}, {Bernard}, {Botticella},
  {Casertano}, {Chastel}, {Chen}, {Chen}, {Cole}, {Deacon}, {Frenk},
  {Fitzsimmons}, {Gezari}, {Goessl}, {Goggia}, {Goldman}, {Grebel}, {Hambly},
  {Hasinger}, {Heavens}, {Heckman}, {Henderson}, {Henning}, {Holman}, {Hopp},
  {Ip}, {Isani}, {Keyes}, {Koekemoer}, {Kotak}, {Long}, {Lucey}, {Liu},
  {Martin}, {McLean}, {Morganson}, {Murphy}, {Nieto-Santisteban}, {Norberg},
  {Peacock}, {Pier}, {Postman}, {Primak}, {Rae}, {Rest}, {Riess}, {Riffeser},
  {Rix}, {Roser}, {Schilbach}, {Schultz}, {Scolnic}, {Szalay}, {Seitz},
  {Shiao}, {Small}, {Smith}, {Soderblom}, {Taylor}, {Thakar}, {Thiel},
  {Thilker}, {Urata}, {Valenti}, {Walter}, {Watters}, {Werner}, {White},
  {Wood-Vasey}, \& {Wyse}}]{2016arXiv161205560C}
{Chambers}, K.~C., {Magnier}, E.~A., {Metcalfe}, N., {et~al.} 2016, ArXiv
  e-prints [\eprint[arXiv]{1612.05560}]

\bibitem[{{Corlies} {et~al.}(2018){Corlies}, {Johnston}, \&
  {Wise}}]{2018MNRAS.475.4868C}
{Corlies}, L., {Johnston}, K.~V., \& {Wise}, J.~H. 2018, \mnras, 475, 4868

\bibitem[{{de los Reyes} \& {Kennicutt}(2019)}]{2019ApJ...872...16D}
{de los Reyes}, M. A.~C. \& {Kennicutt}, Robert~C., J. 2019, \apj, 872, 16

\bibitem[{{DES Collaboration}(2016)}]{2016MNRAS.460.1270D}
{DES Collaboration}. 2016, \mnras, 460, 1270

\bibitem[{{Dopita} \& {Smith}(1986)}]{1986ApJ...304..283D}
{Dopita}, M.~A. \& {Smith}, G.~H. 1986, \apj, 304, 283

\bibitem[{{Emerick} {et~al.}(2019){Emerick}, {Bryan}, \& {Mac
  Low}}]{2019MNRAS.482.1304E}
{Emerick}, A., {Bryan}, G.~L., \& {Mac Low}, M.-M. 2019, \mnras, 482, 1304

\bibitem[{{Emerick} {et~al.}(2016){Emerick}, {Mac Low}, {Grcevich}, \&
  {Gatto}}]{2016ApJ...826..148E}
{Emerick}, A., {Mac Low}, M.-M., {Grcevich}, J., \& {Gatto}, A. 2016, \apj,
  826, 148

\bibitem[{{Faucher-Gigu{\`e}re}(2018)}]{2018NatAs.tmp...32F}
{Faucher-Gigu{\`e}re}, C.-A. 2018, Nature Astronomy

\bibitem[{{Ferguson} \& {Binggeli}(1994)}]{1994A&ARv...6...67F}
{Ferguson}, H.~C. \& {Binggeli}, B. 1994, \aapr, 6, 67

\bibitem[{{Homma} {et~al.}(2016){Homma}, {Chiba}, {Okamoto}, {Komiyama},
  {Tanaka}, {Tanaka}, {Ishigaki}, {Akiyama}, {Arimoto}, {Garmilla}, {Lupton},
  {Strauss}, {Furusawa}, {Miyazaki}, {Murayama}, {Nishizawa}, {Takada},
  {Usuda}, \& {Wang}}]{2016ApJ...832...21H}
{Homma}, D., {Chiba}, M., {Okamoto}, S., {et~al.} 2016, \apj, 832, 21

\bibitem[{{Homma} {et~al.}(2018){Homma}, {Chiba}, {Okamoto}, {Komiyama},
  {Tanaka}, {Tanaka}, {Ishigaki}, {Hayashi}, {Arimoto}, {Garmilla}, {Lupton},
  {Strauss}, {Miyazaki}, {Wang}, \& {Murayama}}]{2018PASJ...70S..18H}
{Homma}, D., {Chiba}, M., {Okamoto}, S., {et~al.} 2018, \pasj, 70, S18

\bibitem[{{Jeon} {et~al.}(2017){Jeon}, {Besla}, \&
  {Bromm}}]{2017ApJ...848...85J}
{Jeon}, M., {Besla}, G., \& {Bromm}, V. 2017, \apj, 848, 85

\bibitem[{{Je{\v r}{\'a}bkov{\'a}} {et~al.}(2018){Je{\v r}{\'a}bkov{\'a}},
  {Hasani Zonoozi}, {Kroupa}, {Beccari}, {Yan}, {Vazdekis}, \&
  {Zhang}}]{2018A&A...620A..39J}
{Je{\v r}{\'a}bkov{\'a}}, T., {Hasani Zonoozi}, A., {Kroupa}, P., {et~al.}
  2018, \aap, 620, A39

\bibitem[{{Katz}(1992)}]{1992ApJ...391..502K}
{Katz}, N. 1992, \apj, 391, 502

\bibitem[{{Kim} \& {Ostriker}(2015)}]{2015ApJ...802...99K}
{Kim}, C.-G. \& {Ostriker}, E.~C. 2015, \apj, 802, 99

\bibitem[{{Klypin} {et~al.}(1999){Klypin}, {Gottl{\"o}ber}, {Kravtsov}, \&
  {Khokhlov}}]{1999ApJ...516..530K}
{Klypin}, A., {Gottl{\"o}ber}, S., {Kravtsov}, A.~V., \& {Khokhlov}, A.~M.
  1999, \apj, 516, 530

\bibitem[{{Komatsu} {et~al.}(2009){Komatsu}, {Dunkley}, {Nolta}, {Bennett},
  {Gold}, {Hinshaw}, {Jarosik}, {Larson}, {Limon}, {Page}, {Spergel},
  {Halpern}, {Hill}, {Kogut}, {Meyer}, {Tucker}, {Weiland}, {Wollack}, \&
  {Wright}}]{2009ApJS..180..330K}
{Komatsu}, E., {Dunkley}, J., {Nolta}, M.~R., {et~al.} 2009, \apjs, 180, 330

\bibitem[{{Krause} {et~al.}(2016){Krause}, {Charbonnel}, {Bastian}, \&
  {Diehl}}]{2016A&A...587A..53K}
{Krause}, M.~G.~H., {Charbonnel}, C., {Bastian}, N., \& {Diehl}, R. 2016, \aap,
  587, A53

\bibitem[{{Kroupa}(2001)}]{2001MNRAS.322..231K}
{Kroupa}, P. 2001, \mnras, 322, 231

\bibitem[{{Lada} \& {Lada}(2003)}]{2003ARA&A..41...57L}
{Lada}, C.~J. \& {Lada}, E.~A. 2003, \araa, 41, 57

\bibitem[{{Laevens} {et~al.}(2015){Laevens}, {Martin}, {Bernard}, {Schlafly},
  {Sesar}, {Rix}, {Bell}, {Ferguson}, {Slater}, {Sweeney}, {Wyse}, {Huxor},
  {Burgett}, {Chambers}, {Draper}, {Hodapp}, {Kaiser}, {Magnier}, {Metcalfe},
  {Tonry}, {Wainscoat}, \& {Waters}}]{2015ApJ...813...44L}
{Laevens}, B.~P.~M., {Martin}, N.~F., {Bernard}, E.~J., {et~al.} 2015, \apj,
  813, 44

\bibitem[{{Larson}(1974)}]{1974MNRAS.169..229L}
{Larson}, R.~B. 1974, \mnras, 169, 229

\bibitem[{{Leitherer} {et~al.}(2014){Leitherer}, {Ekstr{\"o}m}, {Meynet},
  {Schaerer}, {Agienko}, \& {Levesque}}]{2014ApJS..212...14L}
{Leitherer}, C., {Ekstr{\"o}m}, S., {Meynet}, G., {et~al.} 2014, \apjs, 212, 14

\bibitem[{{Mac Low} \& {Ferrara}(1999)}]{1999ApJ...513..142M}
{Mac Low}, M.-M. \& {Ferrara}, A. 1999, \apj, 513, 142

\bibitem[{{Mac Low} \& {McCray}(1988)}]{1988ApJ...324..776M}
{Mac Low}, M.-M. \& {McCray}, R. 1988, \apj, 324, 776

\bibitem[{{Marcolini} {et~al.}(2006){Marcolini}, {D'Ercole}, {Brighenti}, \&
  {Recchi}}]{2006MNRAS.371..643M}
{Marcolini}, A., {D'Ercole}, A., {Brighenti}, F., \& {Recchi}, S. 2006, \mnras,
  371, 643

\bibitem[{{Martizzi} {et~al.}(2015){Martizzi}, {Faucher-Gigu{\`e}re}, \&
  {Quataert}}]{2015MNRAS.450..504M}
{Martizzi}, D., {Faucher-Gigu{\`e}re}, C.-A., \& {Quataert}, E. 2015, \mnras,
  450, 504

\bibitem[{{McKee} \& {Williams}(1997)}]{1997ApJ...476..144M}
{McKee}, C.~F. \& {Williams}, J.~P. 1997, \apj, 476, 144

\bibitem[{{Moore} {et~al.}(1999){Moore}, {Ghigna}, {Governato}, {Lake},
  {Quinn}, {Stadel}, \& {Tozzi}}]{1999ApJ...524L..19M}
{Moore}, B., {Ghigna}, S., {Governato}, F., {et~al.} 1999, \apjl, 524, L19

\bibitem[{{Mori} {et~al.}(2002){Mori}, {Ferrara}, \&
  {Madau}}]{2002ApJ...571...40M}
{Mori}, M., {Ferrara}, A., \& {Madau}, P. 2002, \apj, 571, 40

\bibitem[{{Ni} {et~al.}(2018){Ni}, {Di Matteo}, {Feng}, {Croft}, \&
  {Tenneti}}]{2018MNRAS.481.4877N}
{Ni}, Y., {Di Matteo}, T., {Feng}, Y., {Croft}, R.~A.~C., \& {Tenneti}, A.
  2018, \mnras, 481, 4877

\bibitem[{{Okamoto} {et~al.}(2012){Okamoto}, {Arimoto}, {Yamada}, \&
  {Onodera}}]{2012ApJ...744...96O}
{Okamoto}, S., {Arimoto}, N., {Yamada}, Y., \& {Onodera}, M. 2012, \apj, 744,
  96

\bibitem[{{Plummer}(1911)}]{1911MNRAS..71..460P}
{Plummer}, H.~C. 1911, \mnras, 71, 460

\bibitem[{{Read} {et~al.}(2016){Read}, {Agertz}, \&
  {Collins}}]{2016MNRAS.459.2573R}
{Read}, J.~I., {Agertz}, O., \& {Collins}, M.~L.~M. 2016, \mnras, 459, 2573

\bibitem[{{Recchi} \& {Hensler}(2013)}]{2013A&A...551A..41R}
{Recchi}, S. \& {Hensler}, G. 2013, \aap, 551, A41

\bibitem[{{Roberts}(1957)}]{1957PASP...69...59R}
{Roberts}, M.~S. 1957, \pasp, 69, 59

\bibitem[{{Robles-Valdez} {et~al.}(2017){Robles-Valdez},
  {Rodr{\'{\i}}guez-Gonz{\'a}lez}, {Hern{\'a}ndez-Mart{\'{\i}}nez}, \&
  {Esquivel}}]{2017ApJ...835..136R}
{Robles-Valdez}, F., {Rodr{\'{\i}}guez-Gonz{\'a}lez}, A.,
  {Hern{\'a}ndez-Mart{\'{\i}}nez}, L., \& {Esquivel}, A. 2017, \apj, 835, 136

\bibitem[{{Romano} {et~al.}(2015){Romano}, {Bellazzini}, {Starkenburg}, \&
  {Leaman}}]{2015MNRAS.446.4220R}
{Romano}, D., {Bellazzini}, M., {Starkenburg}, E., \& {Leaman}, R. 2015,
  \mnras, 446, 4220

\bibitem[{{Rosen} \& {Bregman}(1995)}]{1995ApJ...440..634R}
{Rosen}, A. \& {Bregman}, J.~N. 1995, \apj, 440, 634

\bibitem[{{Salvadori} \& {Ferrara}(2009)}]{2009MNRAS.395L...6S}
{Salvadori}, S. \& {Ferrara}, A. 2009, \mnras, 395, L6

\bibitem[{{Sawala} {et~al.}(2016){Sawala}, {Frenk}, {Fattahi}, {Navarro},
  {Theuns}, {Bower}, {Crain}, {Furlong}, {Jenkins}, {Schaller}, \&
  {Schaye}}]{2016MNRAS.456...85S}
{Sawala}, T., {Frenk}, C.~S., {Fattahi}, A., {et~al.} 2016, \mnras, 456, 85

\bibitem[{{Scannapieco} \& {Br{\"u}ggen}(2010)}]{2010MNRAS.405.1634S}
{Scannapieco}, E. \& {Br{\"u}ggen}, M. 2010, \mnras, 405, 1634

\bibitem[{{Schmidt}(1959)}]{1959ApJ...129..243S}
{Schmidt}, M. 1959, \apj, 129, 243

\bibitem[{{Shanks} {et~al.}(2015){Shanks}, {Metcalfe}, {Chehade}, {Findlay},
  {Irwin}, {Gonzalez-Solares}, {Lewis}, {Yoldas}, {Mann}, {Read}, {Sutorius},
  \& {Voutsinas}}]{2015MNRAS.451.4238S}
{Shanks}, T., {Metcalfe}, N., {Chehade}, B., {et~al.} 2015, \mnras, 451, 4238

\bibitem[{{Silich} \& {Tenorio-Tagle}(2001)}]{2001ApJ...552...91S}
{Silich}, S. \& {Tenorio-Tagle}, G. 2001, \apj, 552, 91

\bibitem[{{Simon}(2019)}]{2019arXiv190105465S}
{Simon}, J.~D. 2019, arXiv e-prints [\eprint[arXiv]{1901.05465}]

\bibitem[{{Simpson} {et~al.}(2015){Simpson}, {Bryan}, {Hummels}, \&
  {Ostriker}}]{2015ApJ...809...69S}
{Simpson}, C.~M., {Bryan}, G.~L., {Hummels}, C., \& {Ostriker}, J.~P. 2015,
  \apj, 809, 69

\bibitem[{{Springel} {et~al.}(2001){Springel}, {Yoshida}, \&
  {White}}]{2001NewA....6...79S}
{Springel}, V., {Yoshida}, N., \& {White}, S.~D.~M. 2001, \na, 6, 79

\bibitem[{{Sutherland}(2010)}]{2010Ap&SS.327..173S}
{Sutherland}, R.~S. 2010, \apss, 327, 173

\bibitem[{{Sutherland} \& {Dopita}(1993)}]{1993ApJS...88..253S}
{Sutherland}, R.~S. \& {Dopita}, M.~A. 1993, \apjs, 88, 253

\bibitem[{{Teyssier}(2002)}]{2002A&A...385..337T}
{Teyssier}, R. 2002, \aap, 385, 337

\bibitem[{{Teyssier} {et~al.}(2013){Teyssier}, {Pontzen}, {Dubois}, \&
  {Read}}]{2013MNRAS.429.3068T}
{Teyssier}, R., {Pontzen}, A., {Dubois}, Y., \& {Read}, J.~I. 2013, \mnras,
  429, 3068

\bibitem[{{Thacker} \& {Couchman}(2000)}]{2000ApJ...545..728T}
{Thacker}, R.~J. \& {Couchman}, H.~M.~P. 2000, \apj, 545, 728

\bibitem[{{Tollerud} {et~al.}(2008){Tollerud}, {Bullock}, {Strigari}, \&
  {Willman}}]{2008ApJ...688..277T}
{Tollerud}, E.~J., {Bullock}, J.~S., {Strigari}, L.~E., \& {Willman}, B. 2008,
  \apj, 688, 277

\bibitem[{{Torrealba} {et~al.}(2016){Torrealba}, {Koposov}, {Belokurov},
  {Irwin}, {Collins}, {Spencer}, {Ibata}, {Mateo}, {Bonaca}, \&
  {Jethwa}}]{2016MNRAS.463..712T}
{Torrealba}, G., {Koposov}, S.~E., {Belokurov}, V., {et~al.} 2016, \mnras, 463,
  712

\bibitem[{{Townsley} {et~al.}(2006){Townsley}, {Broos}, {Feigelson}, {Brandl},
  {Chu}, {Garmire}, \& {Pavlov}}]{2006AJ....131.2140T}
{Townsley}, L.~K., {Broos}, P.~S., {Feigelson}, E.~D., {et~al.} 2006, \aj, 131,
  2140

\bibitem[{{Vader}(1986)}]{1986ApJ...305..669V}
{Vader}, J.~P. 1986, \apj, 305, 669

\bibitem[{{Vincenzo} {et~al.}(2014){Vincenzo}, {Matteucci}, {Vattakunnel}, \&
  {Lanfranchi}}]{2014MNRAS.441.2815V}
{Vincenzo}, F., {Matteucci}, F., {Vattakunnel}, S., \& {Lanfranchi}, G.~A.
  2014, \mnras, 441, 2815

\bibitem[{{Vorobyov} {et~al.}(2015){Vorobyov}, {Recchi}, \&
  {Hensler}}]{2015A&A...579A...9V}
{Vorobyov}, E.~I., {Recchi}, S., \& {Hensler}, G. 2015, \aap, 579, A9

\bibitem[{{Webster} {et~al.}(2015){Webster}, {Bland-Hawthorn}, \&
  {Sutherland}}]{2015ApJ...799L..21W}
{Webster}, D., {Bland-Hawthorn}, J., \& {Sutherland}, R. 2015, \apjl, 799, L21

\bibitem[{{Wheeler} {et~al.}(2018){Wheeler}, {Hopkins}, {Pace},
  {Garrison-Kimmel}, {Boylan-Kolchin}, {Wetzel}, {Bullock}, {Keres},
  {Faucher-Giguere}, \& {Quataert}}]{2018arXiv181202749W}
{Wheeler}, C., {Hopkins}, P.~F., {Pace}, A.~B., {et~al.} 2018, arXiv e-prints
  [\eprint[arXiv]{1812.02749}]

\bibitem[{{Wheeler} {et~al.}(2015){Wheeler}, {O{\~n}orbe}, {Bullock},
  {Boylan-Kolchin}, {Elbert}, {Garrison-Kimmel}, {Hopkins}, \& {Kere{\v
  s}}}]{2015MNRAS.453.1305W}
{Wheeler}, C., {O{\~n}orbe}, J., {Bullock}, J.~S., {et~al.} 2015, \mnras, 453,
  1305

\bibitem[{{White} \& {Rees}(1978)}]{1978MNRAS.183..341W}
{White}, S.~D.~M. \& {Rees}, M.~J. 1978, \mnras, 183, 341

\bibitem[{{Willman} {et~al.}(2005){Willman}, {Blanton}, {West}, {Dalcanton},
  {Hogg}, {Schneider}, {Wherry}, {Yanny}, \& {Brinkmann}}]{2005AJ....129.2692W}
{Willman}, B., {Blanton}, M.~R., {West}, A.~A., {et~al.} 2005, \aj, 129, 2692

\bibitem[{{Wise} {et~al.}(2012){Wise}, {Abel}, {Turk}, {Norman}, \&
  {Smith}}]{2012MNRAS.427..311W}
{Wise}, J.~H., {Abel}, T., {Turk}, M.~J., {Norman}, M.~L., \& {Smith}, B.~D.
  2012, \mnras, 427, 311

\bibitem[{{Wolf} {et~al.}(2010){Wolf}, {Martinez}, {Bullock}, {Kaplinghat},
  {Geha}, {Mu{\~n}oz}, {Simon}, \& {Avedo}}]{2010MNRAS.406.1220W}
{Wolf}, J., {Martinez}, G.~D., {Bullock}, J.~S., {et~al.} 2010, \mnras, 406,
  1220

\bibitem[{{Yadav} {et~al.}(2017){Yadav}, {Mukherjee}, {Sharma}, \&
  {Nath}}]{2017MNRAS.465.1720Y}
{Yadav}, N., {Mukherjee}, D., {Sharma}, P., \& {Nath}, B.~B. 2017, \mnras, 465,
  1720

\bibitem[{{Yan} {et~al.}(2017){Yan}, {Jerabkova}, \&
  {Kroupa}}]{2017A&A...607A.126Y}
{Yan}, Z., {Jerabkova}, T., \& {Kroupa}, P. 2017, \aap, 607, A126

\bibitem[{{York} {et~al.}(2000){York}, {Adelman}, {Anderson}, {Anderson},
  {Annis}, {Bahcall}, {Bakken}, {Barkhouser}, {Bastian}, {Berman}, {Boroski},
  {Bracker}, {Briegel}, {Briggs}, {Brinkmann}, {Brunner}, {Burles}, {Carey},
  {Carr}, {Castander}, {Chen}, {Colestock}, {Connolly}, {Crocker}, {Csabai},
  {Czarapata}, {Davis}, {Doi}, {Dombeck}, {Eisenstein}, {Ellman}, {Elms},
  {Evans}, {Fan}, {Federwitz}, {Fiscelli}, {Friedman}, {Frieman}, {Fukugita},
  {Gillespie}, {Gunn}, {Gurbani}, {de Haas}, {Haldeman}, {Harris}, {Hayes},
  {Heckman}, {Hennessy}, {Hindsley}, {Holm}, {Holmgren}, {Huang}, {Hull},
  {Husby}, {Ichikawa}, {Ichikawa}, {Ivezi{\'c}}, {Kent}, {Kim}, {Kinney},
  {Klaene}, {Kleinman}, {Kleinman}, {Knapp}, {Korienek}, {Kron}, {Kunszt},
  {Lamb}, {Lee}, {Leger}, {Limmongkol}, {Lindenmeyer}, {Long}, {Loomis},
  {Loveday}, {Lucinio}, {Lupton}, {MacKinnon}, {Mannery}, {Mantsch}, {Margon},
  {McGehee}, {McKay}, {Meiksin}, {Merelli}, {Monet}, {Munn}, {Narayanan},
  {Nash}, {Neilsen}, {Neswold}, {Newberg}, {Nichol}, {Nicinski}, {Nonino},
  {Okada}, {Okamura}, {Ostriker}, {Owen}, {Pauls}, {Peoples}, {Peterson},
  {Petravick}, {Pier}, {Pope}, {Pordes}, {Prosapio}, {Rechenmacher}, {Quinn},
  {Richards}, {Richmond}, {Rivetta}, {Rockosi}, {Ruthmansdorfer}, {Sandford},
  {Schlegel}, {Schneider}, {Sekiguchi}, {Sergey}, {Shimasaku}, {Siegmund},
  {Smee}, {Smith}, {Snedden}, {Stone}, {Stoughton}, {Strauss}, {Stubbs},
  {SubbaRao}, {Szalay}, {Szapudi}, {Szokoly}, {Thakar}, {Tremonti}, {Tucker},
  {Uomoto}, {Vanden Berk}, {Vogeley}, {Waddell}, {Wang}, {Watanabe},
  {Weinberg}, {Yanny}, {Yasuda}, \& {SDSS Collaboration}}]{2000AJ....120.1579Y}
{York}, D.~G., {Adelman}, J., {Anderson}, Jr., J.~E., {et~al.} 2000, \aj, 120,
  1579

\end{thebibliography}

\begin{appendix}

\section{On the distribution of OB associations and convergence test}
\label{sec:AppA}

   \begin{table*}
   \caption{Characteristics of OB associations for different simulations}
   \label{tab:sims}
   \centering
   \begin{tabular}{@{}c@{\hspace{0.2in}} c@{\hspace{0.2in}} l@{}}
   \hline\hline
   Simulation & $\mathscr{N}_{\rm OB}$ & $(r_1,N_1), (r_2,N_2)$ \ldots $(r_{\mathscr{N}_{\rm OB}},N_{\mathscr{N}_{\rm OB}})$ \\
   \hline
   A-10-04-01-P & 10 & (92, 30), (255, 80), (40, 33), (200, 38), (113, 39), (341, 152), (135, 49), (191, 34), (368, 135), (351, 60) \\
   R-10-04-01-P & 10 & (92, 30), (255, 80), (40, 33), (200, 38), (113, 39), (341, 152), (135, 49), (191, 34), (368, 135), (351, 60) \\
   A-10-20-04-P & 10 & (92, 30), (255, 80), (40, 33), (200, 38), (113, 39), (341, 152), (135, 49), (191, 34), (368, 135), (351, 60) \\
   A-07-20-04-P &  7 & (100, 33), (239, 322), (255, 33), (780, 53), (206, 59), (173, 86), (45, 64) \\
   A-04-20-04-P &  4 & (44, 33), (104, 327), (413, 62), (26, 228) \\
   A-04-20-04-C &  4 & (15, 33), (35, 327), (152, 62), (9, 228) \\
   \hline
   \end{tabular}
   \end{table*}

   In order to assess the ability of stellar feedback to lift the gas left over 
   from the star formation process out of the potential well of an isolated, 
   low-mass dark matter halo, $M_{\rm{DM}}$~= 3.5~$\times$ 10$^7$~M$_\odot$, we 
   adopt an idealized set-up.

   Massive stars are assumed to be coeval (i.e., born on a timescale that is 
   short compared to their lifetimes) and grouped in OB associations. It has 
   long been recognized, in fact, that most stars originate in associations 
   \citep{1957PASP...69...59R,2003ARA&A..41...57L}. The probability for an OB 
   association to contain $N$ SNe is set to $f(N) \propto N^{-2}$ 
   \citep[see][and references therein]{1997ApJ...476..144M,
     2015ApJ...814L..14C}, with $30 \le N \le 650$ (the upper limit here is 
   dictated by the total number of SNe that are expected in the system under 
   scrutiny; see Section~\ref{sec:setup}). By applying a grouping procedure and 
   using the above power-law distribution, we end up with a number 
   $\mathscr{N}_{\rm OB}$ of associations, each containing a variable number of 
   massive stars such that the total number of SN progenitors comes to 650.

\begin{figure}
   \centering
   \includegraphics{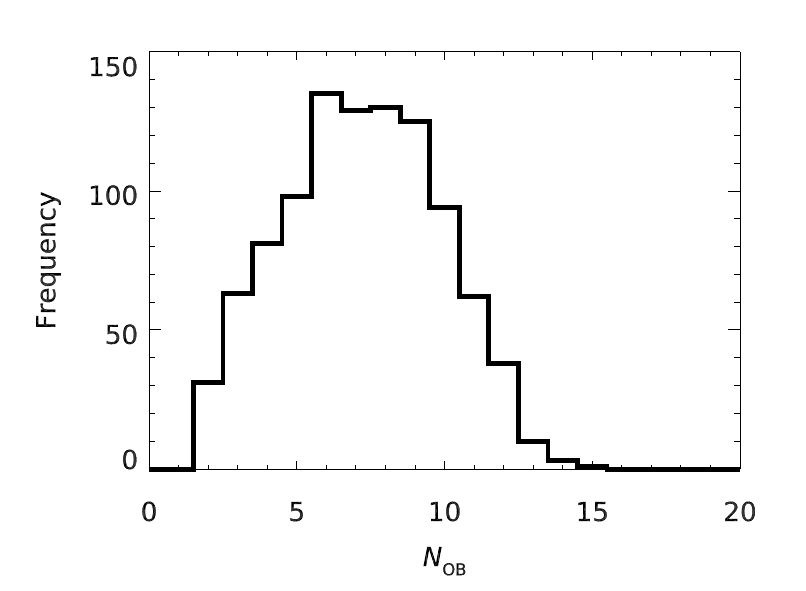}
   \caption{ Occurrence rate of configurations with 650 SNe distributed in 
     $\mathscr{N}_{\rm OB}$ associations, for 1000 random realizations.}
   \label{fig:NOBdistr}
   \end{figure}

\begin{figure}
   \centering
   \includegraphics[width=7.5cm]{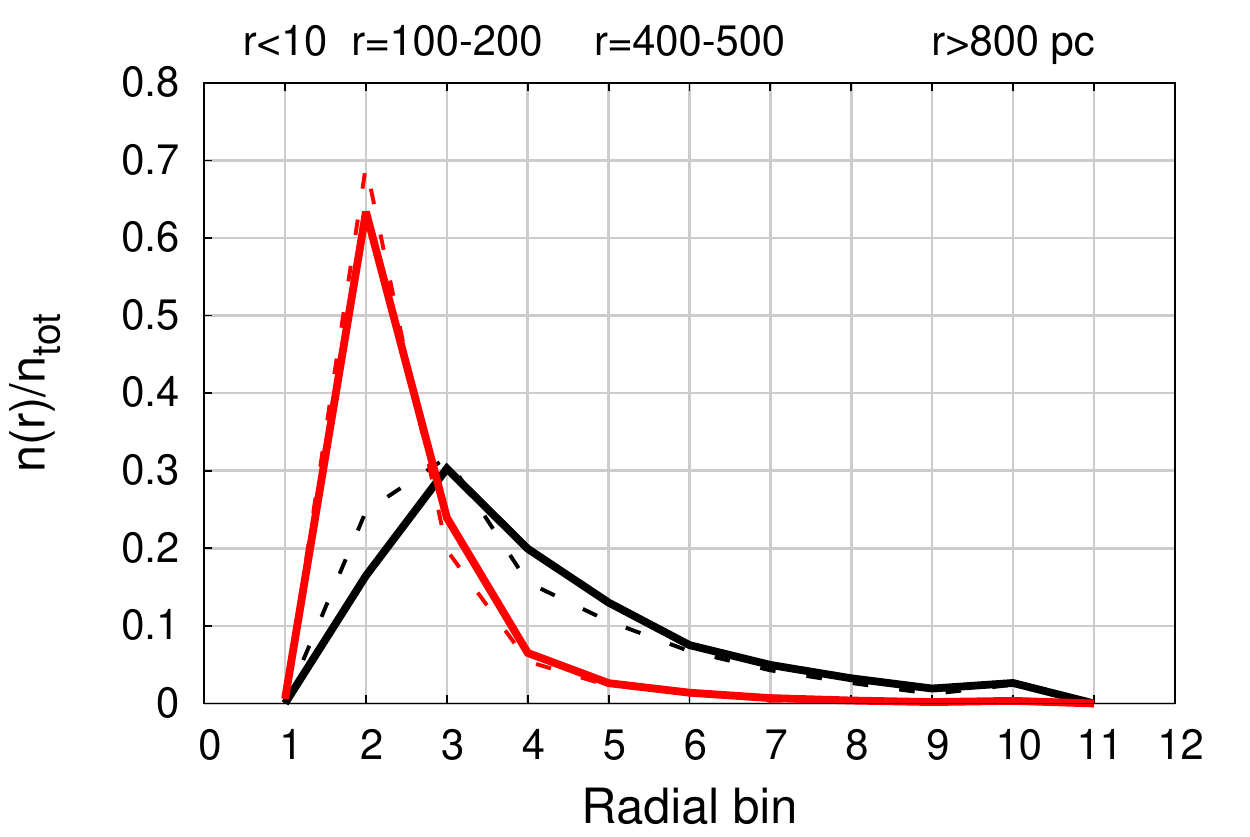}
   \caption{ Relative numbers of OB associations in different radial bins, for 
     1000 random placements of 4 (dashed lines) and 10 (solid lines) 
     associations following a Plummer density profile with either $a \simeq$ 
     200 (black lines) or 60~pc (red lines).}
   \label{fig:rOBdistr}
   \end{figure}

   Fig.~\ref{fig:NOBdistr} shows the frequency distribution of 
   $\mathscr{N}_{\rm OB}$ for 1000 random realizations. The distribution has a 
   peak around $\mathscr{N}_{\rm OB} \simeq$ 6--9, with values lower than 3 and 
   higher than 11 largely disadvantaged. We then need to locate spatially the 
   $\mathscr{N}_{\rm OB}$ associations, which is clearly an extremely noisy 
   procedure. Fig.~\ref{fig:rOBdistr} shows the relative numbers of OB 
   associations that fall in ten galactocentric distance bins ($r <$ 10~pc, 
   $r =$ 10-100~pc, $r =$ 100-200~pc, $r =$ 200-300~pc, $r =$ 300-400~pc, 
   \ldots, $r =$ 700-800~pc, $r >$ 800~pc) after 1000 random placements of 
   $\mathscr{N}_{\rm OB} =$ 4 (dashed lines) or 10 (solid lines) associations 
   drawn from a Plummer density profile with characteristic radius $a \simeq$ 
   200~pc (black lines) or sensibly (70 percent) lower (red lines). The 
   distribution depends strongly on the assumed Plummer radius, with 
   second-order effects due to the actual associations number.

   In the high-resolution simulations discussed in Section~\ref{sec:res}, ten 
   OB associations are disseminated randomly across the simulation volume, 
   according to the same smooth, low-density, single-phase profile 
   \citep[][with $a \simeq$ 200~pc]{1911MNRAS..71..460P} used to describe the 
   gas distribution. With these assumptions, some SN progenitors turn out to be 
   placed in very low-density regions (see Fig.~\ref{fig:densprof}), but we 
   note that there is observational evidence that some extreme systems may form 
   stars notwithstanding their utterly low densities 
   \citep{2018MNRAS.476.4565B}. OB associations located away of the bulk of the 
   ambient medium will be clearly unable to affect it. On the other hand, the 
   denser the surrounding gas, the less efficient the stellar feedback. In 
   order to ascertain the dependence of the overall results and conclusions 
   reported in this paper on the particular set-up choice, we perform a series 
   of lower resolution adiabatic simulations where the number and location of 
   OB associations are let to vary. In particular, in one simulation we suppose 
   that the OB associations form closer to the halo centre ($a \simeq$ 60~pc) 
   from the early collapse and fragmentation of primordial molecular clouds 
   \citep[][]{2000ApJ...540...39A}. We do not investigate more compact 
   configurations, since this would essentially results in handling a 
   proto-globular cluster \citep[see][]{2015ApJ...814L..14C}.

   In Table~\ref{tab:sims} we specify the number and location of the OB 
   associations for the different simulations discussed in this paper. Also 
   given is the number of SNe for each association. A particular simulation is 
   identified with the notation X-yy-zz-ww-V, where X indicates if the 
   simulation is adiabatic (A) or radiative (R), yy is the number of OB 
   associations, zz refers to their size (in pc), ww is the maximum spatial 
   resolution (in pc) and V tells if the associations are drawn from the same, 
   shallow profile used for the gas (P) or if they are more concentrated in the 
   central regions (C).

\begin{figure}
   \centering
   \includegraphics{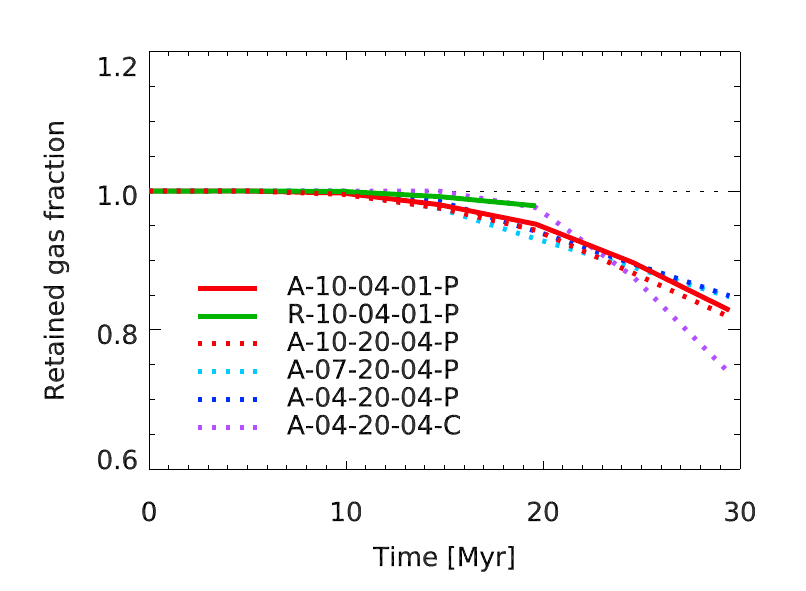}
   \caption{ Evolution of the gas fraction retained by the system (normalized 
     to the initial gas mass), for different resolutions and OB association 
     distributions (see bottom-left corner and Table~\ref{tab:sims}).}
   \label{fig:gascvg}
   \end{figure}

\begin{figure}
   \centering
   \includegraphics{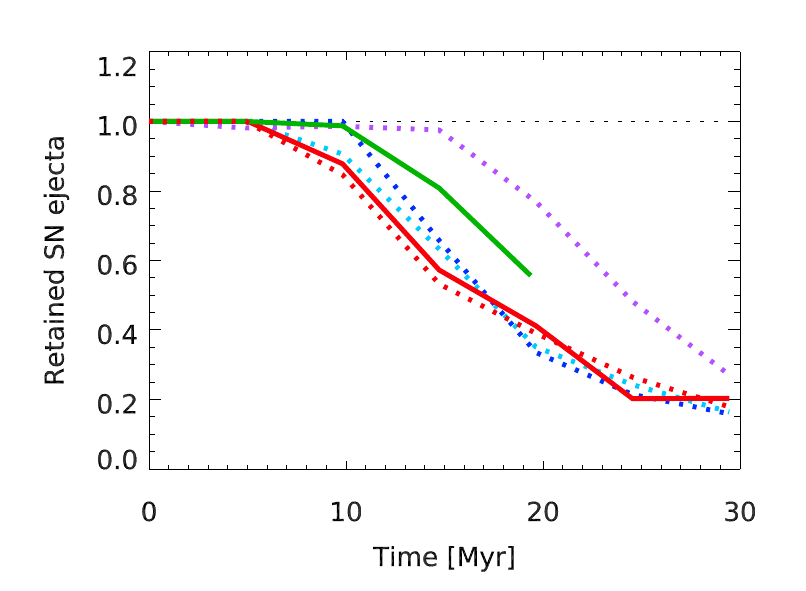}
   \caption{ Same as Fig.~\ref{fig:gascvg}, for the SN ejecta (normalized to 
     the expected theoretical value at each time according to 
     \citealt{2014ApJS..212...14L}).}
   \label{fig:ejecvg}
   \end{figure}

   The results are summarized in Figs.~\ref{fig:gascvg} and \ref{fig:ejecvg}, 
   where we show the fractions of the initial gaseous mass and of the SN ejecta 
   that are retained by the model UFD in the course of different simulations. 
   The quantities from the high-resolution runs are displayed as solid lines, 
   while those from the lower resolution numerical experiments show up as 
   dotted lines. Notwithstanding the convergence criteria by \citet[][see 
     Section~\ref{sec:setup}]{2015ApJ...802...99K} are only partially met, the 
   curves for the lower resolution run A-10-20-04-P track those referring to 
   the corresponding high-resolution simulation A-10-04-01-P pretty well (cfr. 
   the red dotted and red solid lines in Figs.~\ref{fig:gascvg} and 
   \ref{fig:ejecvg}, respectively).

   Overall, it appears that both changing the number of the OB associations and 
   their position within the simulation volume does not change our main 
   conclusion that stellar feedback alone does not suffice to vent the bulk of 
   the neutral ISM out of the model galaxy. However, it must be stressed that 
   we adopt an idealized set-up, which lacks of a structured, multiphase ISM. 
   Furthermore, we consider a stellar population already in place at the 
   beginning of the simulation. In principle, however, according to the IGIMF 
   theory a prolonged star formation in UFDs should result in less massive 
   stars available to heat the gas \citep[][]{2017A&A...607A.126Y,
     2018A&A...620A..39J}, leading to a reduced feedback. Yet, before jumping 
   to conclusions it is mandatory to include a self-consistent recipe for star 
   formation in the simulations, and this will be part of a forthcoming work.

\section{Three-dimensional visualization}
\label{sec:AppB}

   \begin{figure*}
   \centering
   \includegraphics[width=0.41\textwidth]{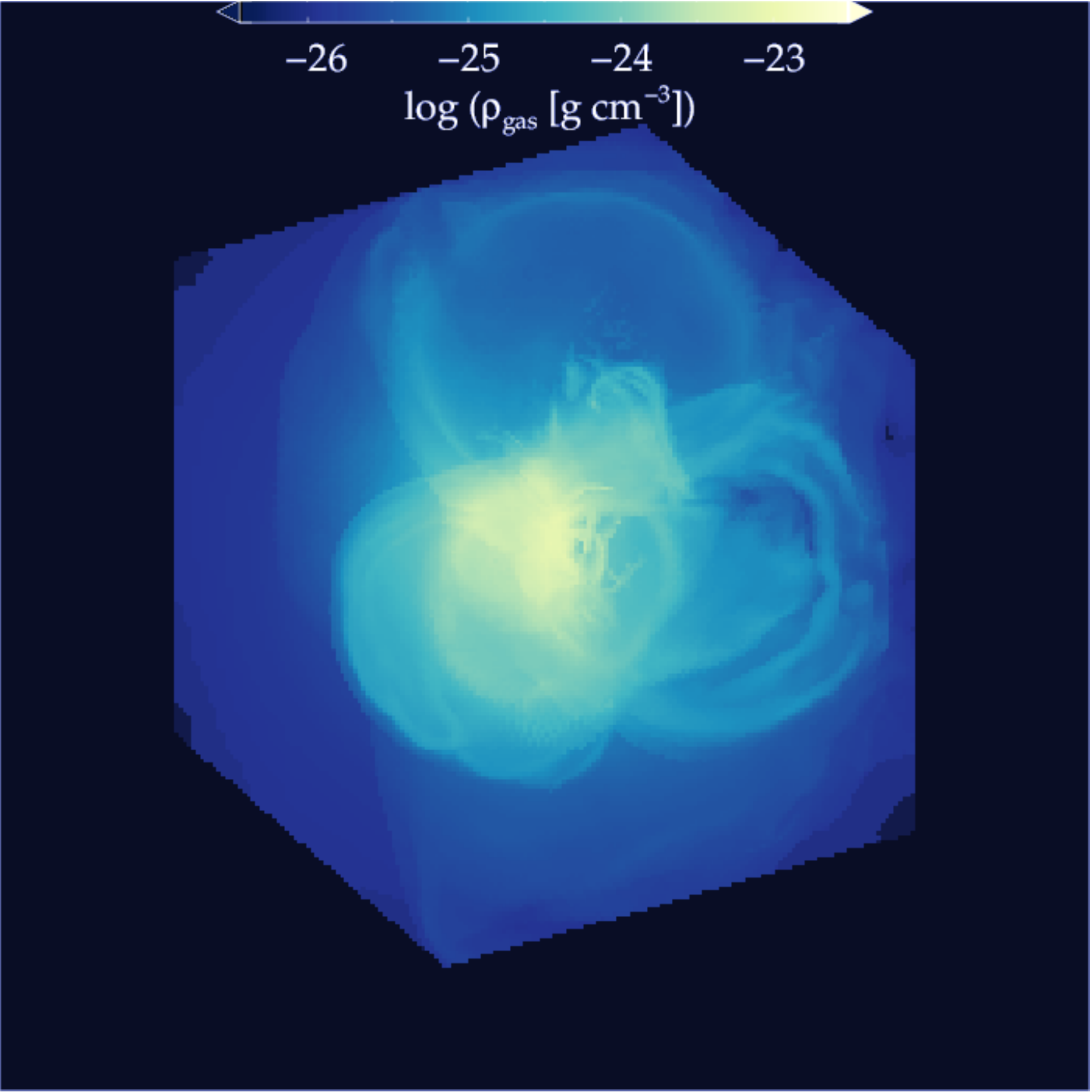}
   \includegraphics[width=0.41\textwidth]{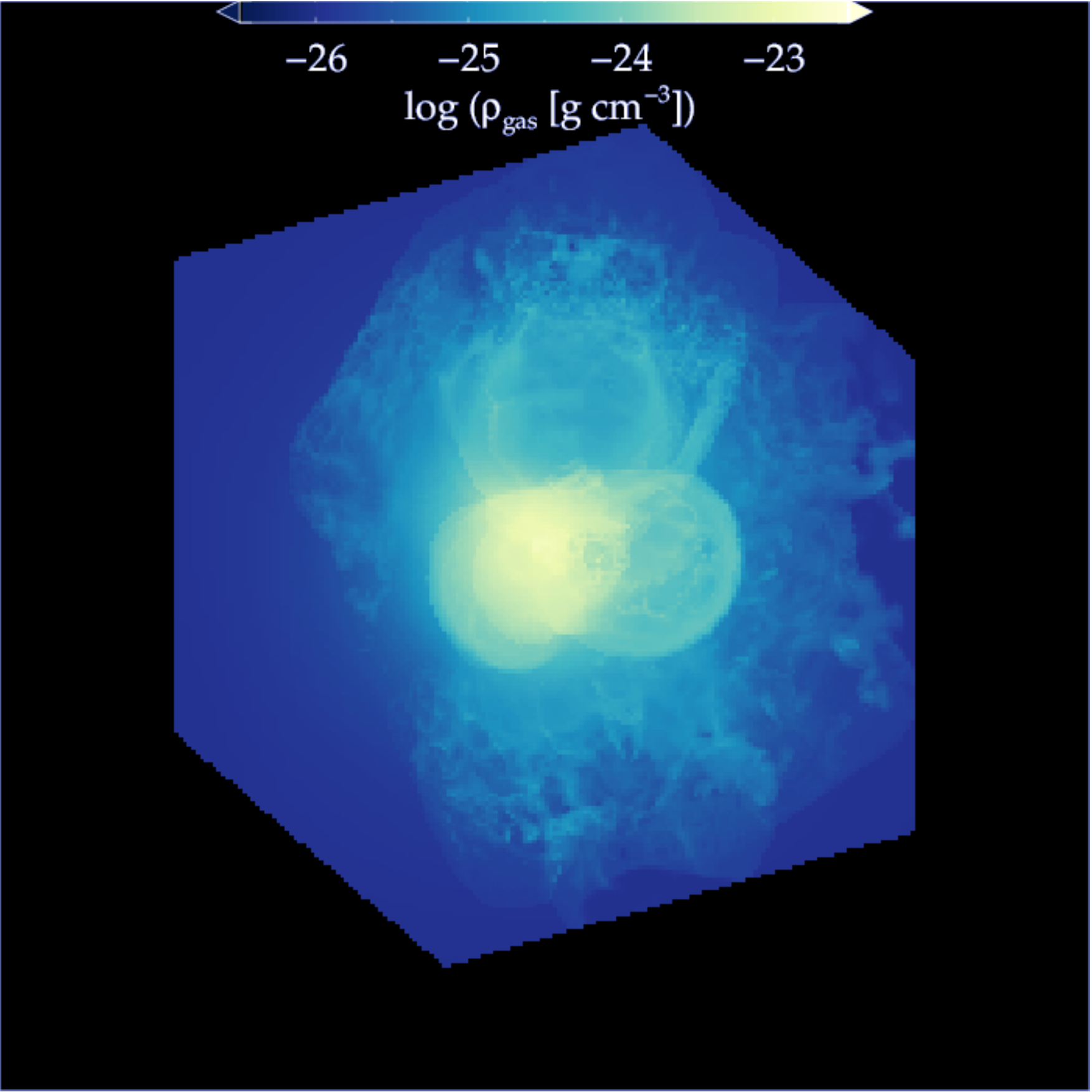}

   \vspace{0.1cm}

   \includegraphics[width=0.41\textwidth]{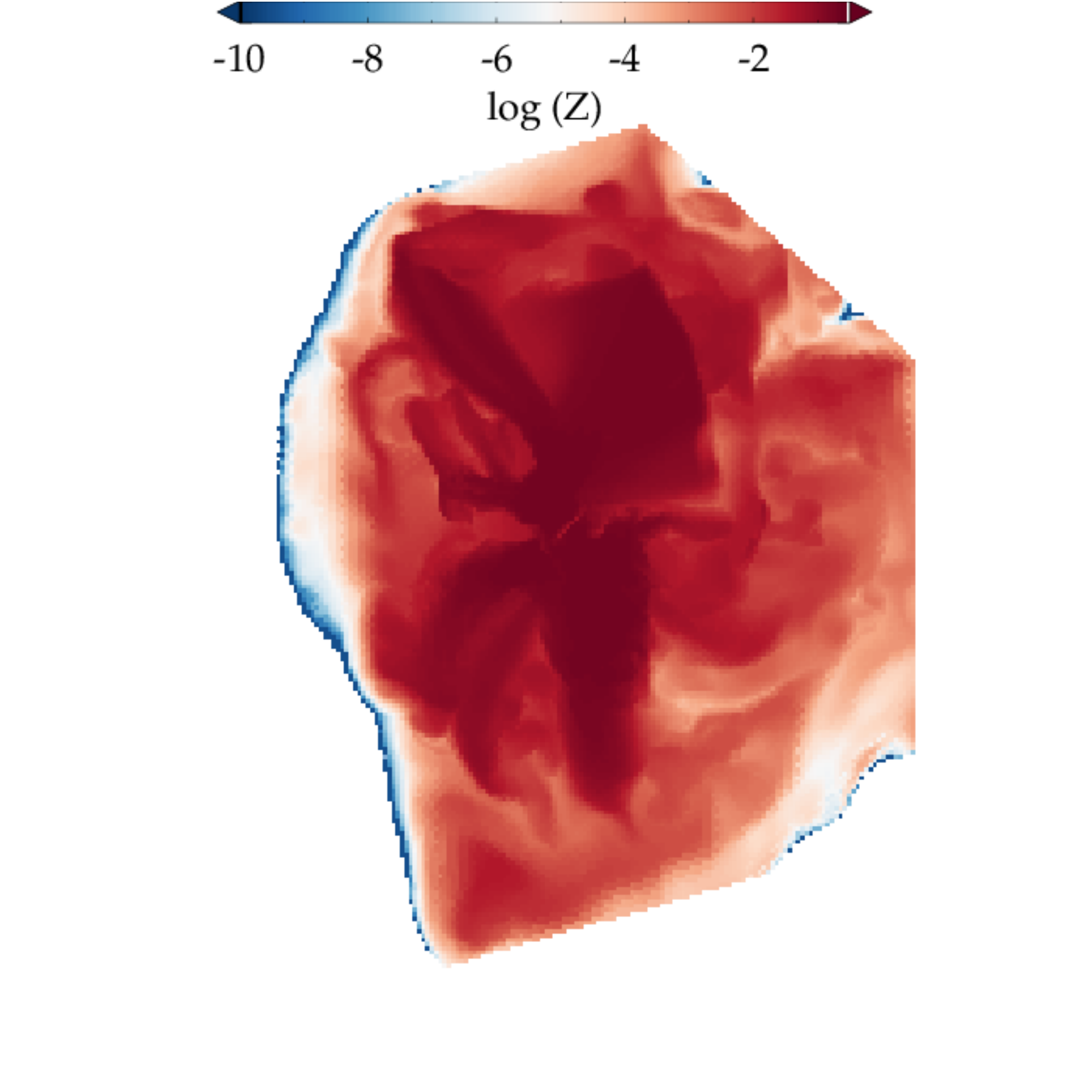}
   \includegraphics[width=0.41\textwidth]{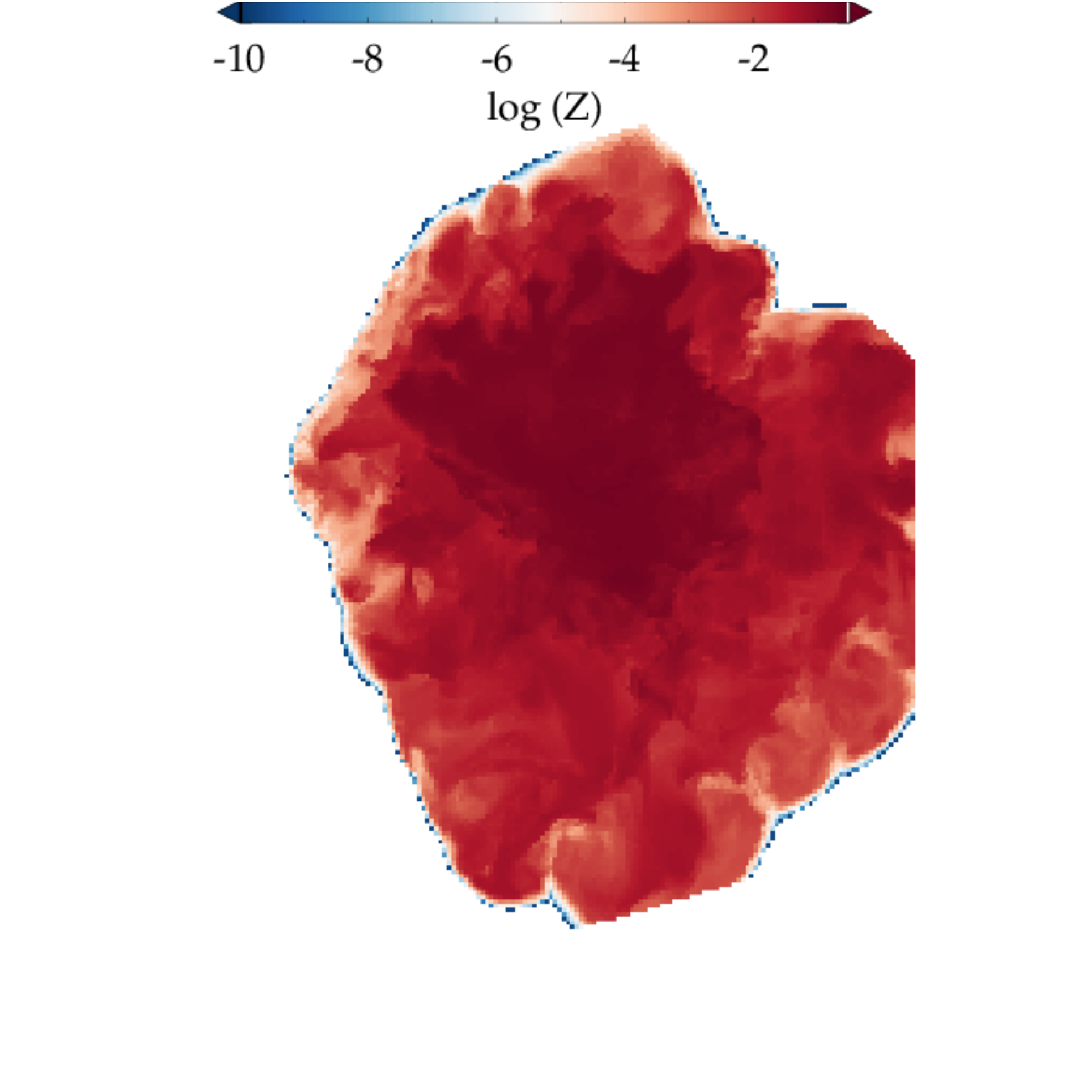}
   \caption{ 3D structure of gas (top images) and metals (bottom images) in the 
     simulation box at $t = 20$~Myr. The images on the left refer to the 
     high-resolution adiabatic run, while those on the right are for the run 
     with cooling.}
   \label{fig:3d20}
   \end{figure*}

   \begin{figure*}
   \centering
   \includegraphics[width=0.41\textwidth]{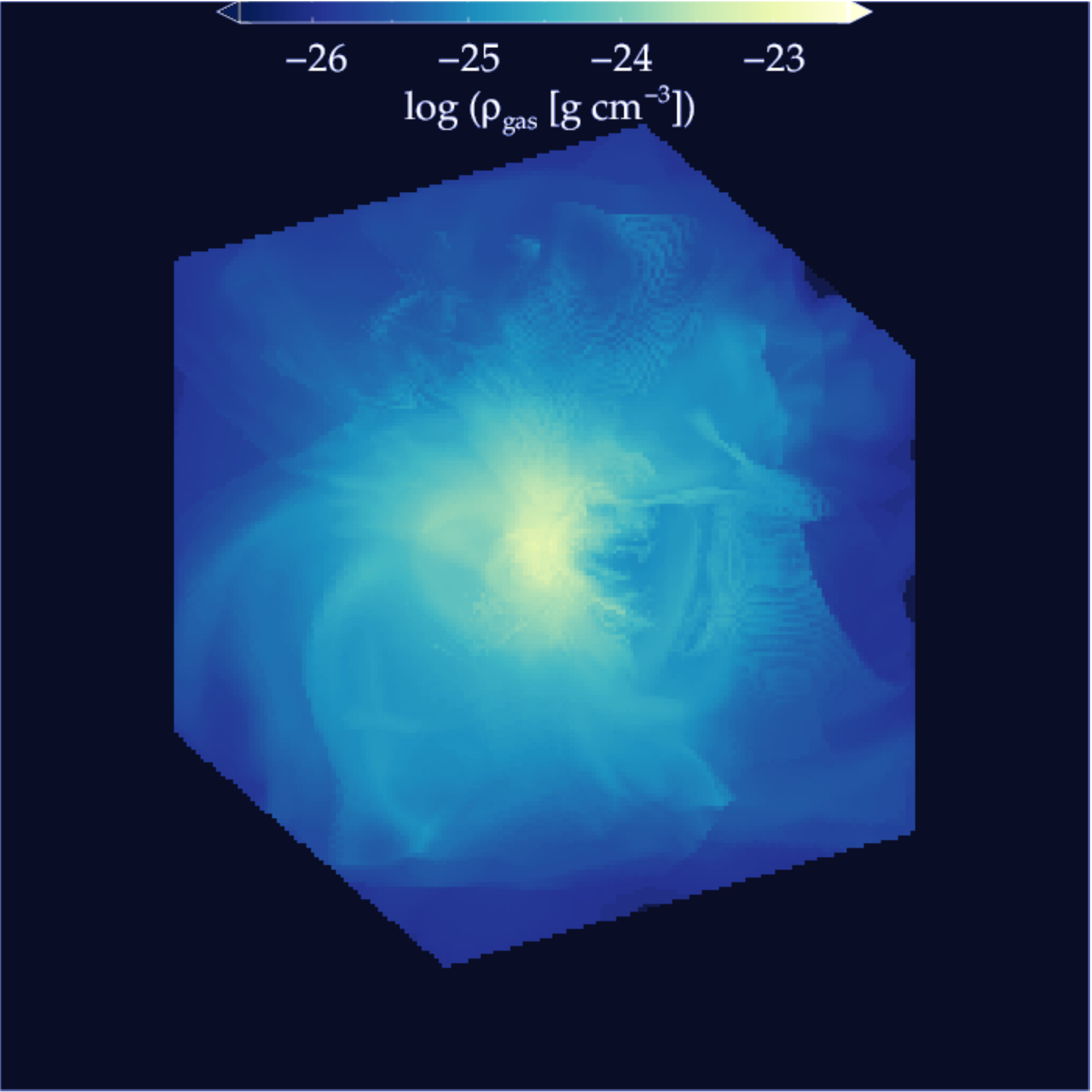}
   \includegraphics[width=0.41\textwidth]{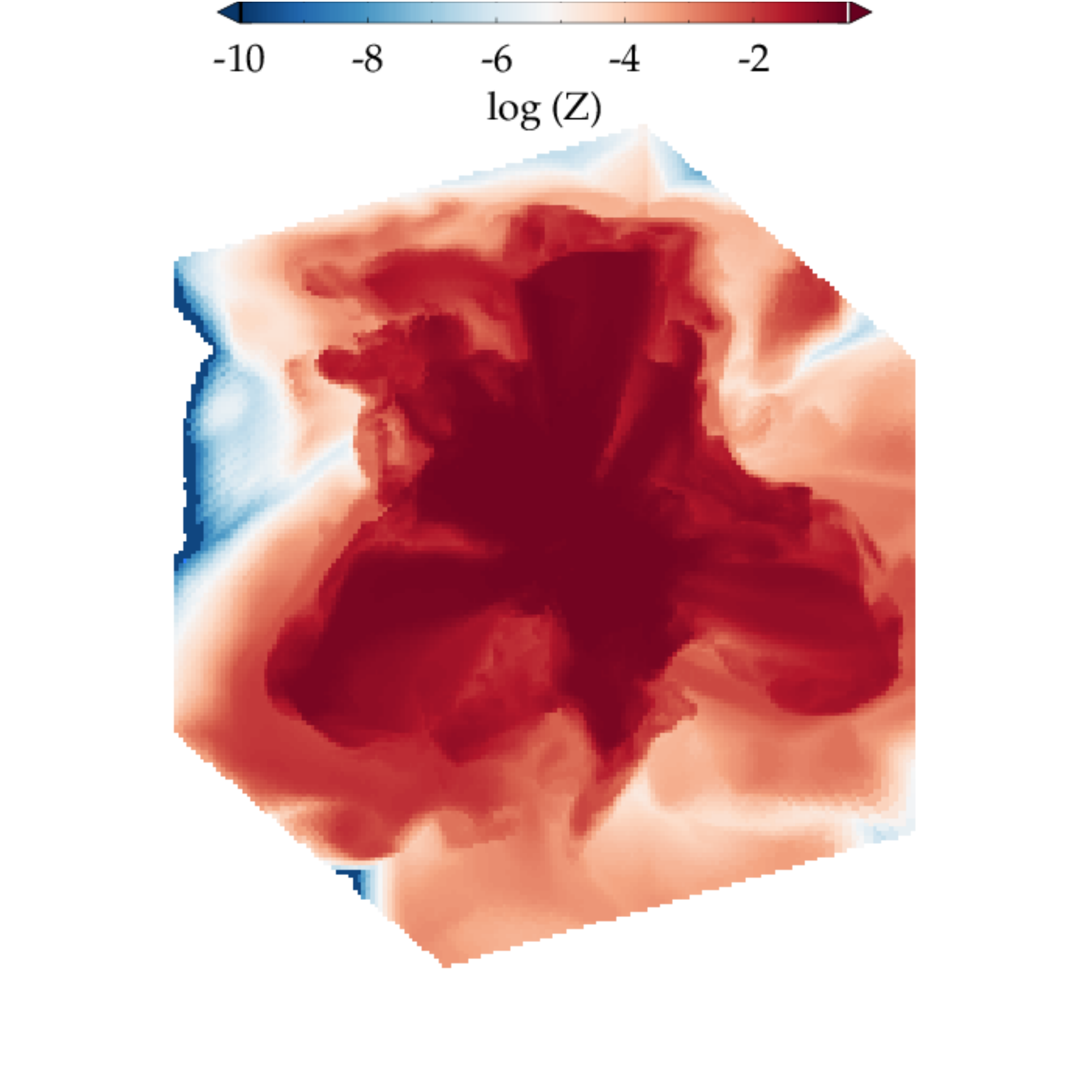}
   \caption{ 3D structure of gas (left-hand image) and metals (right-hand 
     image) in the simulation box at $t = 30$~Myr, for the adiabatic run.}
   \label{fig:3d30}
   \end{figure*}

   Fig.~\ref{fig:3d20} and ~\ref{fig:3d30} display the projections of the 3D 
   distributions of gas and metallicity within the simulation box, at two 
   different times, for our high-resolution simulations. Translucent rendering 
   of the volume is performed by using IDL post-processing utils distributed 
   with the source code.

\end{appendix}

\end{document}